\documentclass[lettersize,journal]{IEEEtran}
\usepackage{amsmath,amsfonts}
\usepackage{algorithmic}
\usepackage{algorithm}
\usepackage{array}
\usepackage[caption=false,font=normalsize,labelfont=sf,textfont=sf]{subfig}
\usepackage{textcomp}
\usepackage{stfloats}
\usepackage{url}
\usepackage{verbatim}
\usepackage{graphicx}
\usepackage{cite}
\usepackage{multirow}%
\usepackage{amssymb}%
\usepackage{amsthm}%
\usepackage{mathrsfs}%
\usepackage{xcolor}%
\usepackage{manyfoot}%
\usepackage{booktabs}%
\usepackage{listings}%
\usepackage{color,xcolor}
\usepackage{colortbl}
\usepackage{makecell,array}

\definecolor{cvprblue}{rgb}{0.21,0.49,0.74}
\definecolor{textpurple}{RGB}{85,100,175}
\definecolor{textgreen}{RGB}{80,180,163}

\definecolor{tablelightblue}{RGB}{227,243,254}
\definecolor{tableblue}{RGB}{226,233,254}
\definecolor{tablegray}{RGB}{242,242,242}

\definecolor{textblue}{RGB}{90,112,164}
\definecolor{textgray}{RGB}{127,127,127}

\newcommand{\gainblueup}[1]{$_{\textcolor{textblue}{\tiny \uparrow #1}}$}
\newcommand{\gainbluedown}[1]{$_{\textcolor{textblue}{\tiny \downarrow #1}}$}

\newcommand{\gaingraydown}[1]{$_{\textcolor{textgray}{\tiny \downarrow #1}}$}

\newcommand{\stdblue}[1]{$_{\textcolor{textblue}{\tiny \pm #1}}$}
\newcommand{\stdpurple}[1]{$_{\textcolor{textpurple}{\tiny \pm #1}}$}
\newcommand{\stdgray}[1]{$_{\textcolor{textgray}{\tiny \pm #1}}$}

\definecolor{ctbg}{RGB}{233,244,240}
\definecolor{mribg}{RGB}{230,237,246}
\definecolor{petbg}{RGB}{238,236,247}

\definecolor{cttxt}{RGB}{0,102,92}
\definecolor{mritxt}{RGB}{38,84,124}
\definecolor{pettxt}{RGB}{98,76,135}

\definecolor{CTGreen}{HTML}{2E8B57} 
\definecolor{MRIBlue}{HTML}{1F77B4} 
\definecolor{PETPurple}{HTML}{7B3294} 
\newif\ifmodalitycolors 
\modalitycolorstrue 
\newcommand{\modCT}{\ifmodalitycolors\textcolor{CTGreen}{\textbf{CT}}\else CT\fi} 
\newcommand{\modMRI}{\ifmodalitycolors\textcolor{MRIBlue}{\textbf{MRI}}\else MRI\fi} 
\newcommand{\modPET}{\ifmodalitycolors\textcolor{PETPurple}{\textbf{PET}}\else PET\fi}

\usepackage[colorlinks=true,
            citecolor=blue,
            linkcolor=blue,
            urlcolor=blue]{hyperref}

\begin{document}

\bstctlcite{IEEEexample:BSTcontrol}

\title{MeDUET: Disentangled Unified Pretraining for 3D Medical Image Synthesis and Analysis}

\author{Junkai Liu, Ling Shao,~\IEEEmembership{Fellow,~IEEE}, 
Le Zhang
        % <-this % stops a space
\thanks{The computations described in this research were performed using the Baskerville Tier 2 HPC service. Baskerville was funded by the EPSRC and UKRI through the World Class Labs scheme (EP/T022221/1) and the Digital Research Infrastructure programme (EP/W032244/1) and is operated by Advanced Research Computing at the University of Birmingham.}% <-this % stops a space
\thanks{Junkai Liu and Le Zhang are with School of Engineering, College of Engineering and Physical Sciences, University of Birmingham, Birmingham, England, United Kingdom (e-mail: jxl1920@student.bham.ac.uk; l.zhang.16@bham.ac.uk).}
\thanks{Ling Shao is with the UCAS-Terminus AI Laboratory, University of Chinese Academy of Sciences, Beijing 100049, China (e-mail: ling.shao@ieee.org).}
}

\markboth{Journal of \LaTeX\ Class Files,~Vol.~14, No.~8, August~2021}%
{Shell \MakeLowercase{\textit{et al.}}: A Sample Article Using IEEEtran.cls for IEEE Journals}


\maketitle

% \begin{abstract}
% This document describes the most common article elements and how to use the IEEEtran class with \LaTeX \ to produce files that are suitable for submission to the IEEE.  IEEEtran can produce conference, journal, and technical note (correspondence) papers with a suitable choice of class options. 
% \end{abstract}

% \begin{IEEEkeywords}
% Article submission, IEEE, IEEEtran, journal, \LaTeX, paper, template, typesetting.
% \end{IEEEkeywords}

\begin{abstract}
Self-supervised learning (SSL) and diffusion models have respectively advanced representation learning and generative modeling for high-dimensional 3D visual data, yet they are often developed as separate paradigms. Their unification remains challenging under multi-source heterogeneity, as anatomical content must be preserved for analysis while acquisition-related style varies across centers and affects synthesis. In this paper, we propose MeDUET, a 3D \textbf{Me}dical image \textbf{D}isentangled \textbf{U}nifi\textbf{E}d Pre\textbf{T}raining framework in the variational autoencoder latent space. MeDUET formulates unified pretraining as an empirical factor identifiability problem, aiming to learn domain-invariant content factors for anatomy and domain-specific style factors for appearance. To improve factor separation, MeDUET first uses token demixing with a standard adversarial domain regularizer to establish basic content-style specialization, and further introduces Mixed Factor Token Distillation and Swap-invariance Quadruplet Contrast to reduce mixed-region factor leakage and organize factor spaces with factor-wise invariance and discriminability. With these learned factors, MeDUET transfers effectively to both synthesis and analysis, yielding higher fidelity, faster convergence, and better controllability for synthesis, while achieving competitive or superior domain generalization and label efficiency on diverse datasets, tasks, and modalities. Overall, MeDUET shows that multi-source heterogeneity can serve as useful supervision, with disentanglement providing an effective interface for unifying 3D medical image synthesis and analysis. Our code is available at \url{https://github.com/JK-Liu7/MeDUET}.

\end{abstract}


% \keywords{Medical Image Synthesis, Medical Image Analysis, Diffusion Models, Self-supervised Learning}

\section{Introduction}
\label{sec:intro}

\IEEEPARstart{R}{ecently}, self-supervised learning (SSL) and diffusion models have respectively advanced representation learning and image 
generation~\cite{Caron_2021_ICCV, He_2022_CVPR, NEURIPS2020_4c5bcfec, Rombach_2022_CVPR, 10938258}, with emerging evidence that diffusion models can further support visual understanding~\cite{Wei_2023_ICCV, Xiang_2023_ICCV}. In medical AI, both paradigms have shown strong promise for 3D image analysis~\cite{Wu_2024_CVPR} and 
synthesis~\cite{11063450}, yet they remain largely separate. Unifying them is appealing because generative modeling can enrich representations with structural and appearance priors, while anatomical semantics can provide more precise conditioning for synthesis, offering bidirectional benefits that a separated pipeline cannot achieve. This motivates our central question: \textit{How can we establish a unified pretraining framework that benefits both 3D medical image synthesis and analysis?}

% Recently, self-supervised learning (SSL) under the pretraining-finetuning paradigm has become a powerful approach for a wide range of tasks~\cite{Caron_2021_ICCV, oquab2024dinov, He_2020_CVPR, He_2022_CVPR}. Meanwhile, diffusion models have achieved remarkable progress in image generation~\cite{NEURIPS2020_4c5bcfec, NEURIPS2021_49ad23d1, Rombach_2022_CVPR}. Motivated by the idea that generative modeling can also support visual understanding, prior studies~\cite{Wei_2023_ICCV, Xiang_2023_ICCV} have shown that diffusion models can learn representations for perception tasks. In medical AI, SSL and generative models have likewise shown great promise for 3D medical image analysis~\cite{11122336, Wu_2024_CVPR} and synthesis~\cite{hamamci2024generatect, 11063450}, respectively. However, the potential of diffusion models for medical perception, namely unified pretraining for both generation and understanding, remains largely underexplored. Such a unified framework is attractive because generative modeling can enrich representation learning with structural and appearance priors, while analysis-oriented anatomical representations can provide more precise guidance for synthesis, offering bidirectional benefits. Inspired by prior work~\cite{Chu_2025_ICCV}, we ask the following research question: \textit{How can we establish a unified pretraining framework that benefits both 3D medical image synthesis and analysis tasks?}

\begin{figure}[t]
  \centering
  \includegraphics[width=\linewidth]{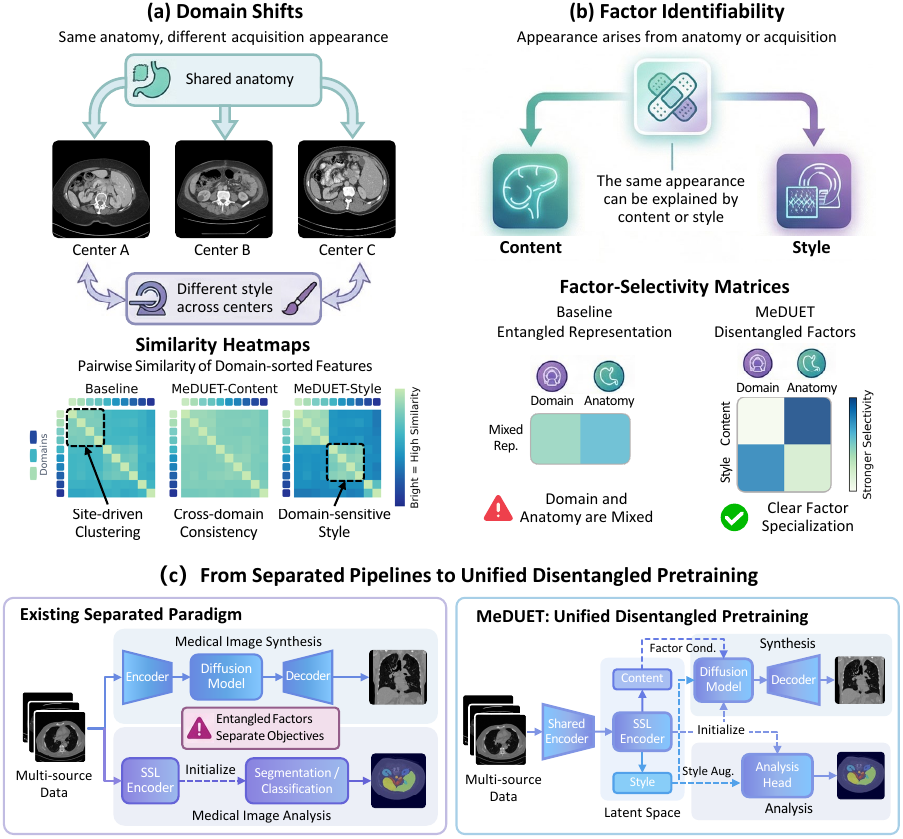}
  \caption{
  (a) Across centers, medical images share similar anatomy but exhibit distinct acquisition styles, leading to domain shifts mainly at the appearance level. 
  (b) Without explicit disentanglement, anatomy and domain information are mixed, whereas MeDUET separates them into empirically disentangled content and style factors. 
  (c) Unlike conventional pipelines that treat synthesis and analysis separately, MeDUET performs unified disentangled pretraining, enabling controllable synthesis and robust analysis.}
  \label{teaser}
  \vspace{-6pt}
\end{figure}

Real-world 3D medical datasets aggregate heterogeneous sites, scanners, field strengths, sequences, and cohorts~\cite{Wu_2024_CVPR, 11274411}, amplifying \textbf{domain shifts} mainly at the \textit{style} level. For example, a liver CT model trained on Hospital A (Siemens, 120 kVp, venous, 2.5 mm, soft kernel) degrades on Hospital B (GE, 100 kVp, arterial, 1.0 mm, sharp) because HU histograms, noise, and edge profiles change while anatomy does not. In contrast, downstream tasks rely on \textit{content} such as organ topology, lesion morphology, and anatomical continuity~\cite{LIU2022102516}. A naïve unification of synthesis and analysis entangles style and anatomy, reducing generator controllability and causing perception models to overfit to style cues, which weakens generalization and label efficiency~\cite{MULLER2025103628}. Thus, domain generalization is vital for unified pretraining but remains underexplored in SSL frameworks (Fig.~\ref{teaser}(a))~\cite{zhang2026unix, TANG2026103770, Wald_2025_CVPR}. 

In light of this challenge, we formalize each volume as composed of two factors, a domain-invariant \textit{content} factor that captures intrinsic anatomy and stable semantics across domains, and a domain-specific \textit{style} factor that reflects acquisition conditions and visual appearance~\cite{MULLER2025103628}. However, learning this decomposition is non-trivial. First, the model receives no explicit supervision specifying which variations belong to anatomy and which to acquisition appearance. Second, clinical style differences are often subtle, where the same anatomy may appear differently across centers while its geometry remains unchanged, so that competing content and style explanations can fit the same observation equally well. As illustrated in Fig.~\ref{teaser}(b), this leads to a \textbf{factor identifiability} challenge~\cite{LIU2022102516, 10634512}. For instance, hyperattenuation in contrast-enhanced CT may arise from true lesion enhancement (content) or from contrast timing and reconstruction kernel (style). Without resolving this ambiguity, entangled factors reduce generator controllability for synthesis while causing analysis models to overfit to domain-specific appearance.

% In light of this challenge and opportunity, we formalize each volume as composed of two factors, a domain-invariant \textit{content} factor that captures intrinsic anatomy and stable semantics across domains, and a domain-specific \textit{style} factor that reflects acquisition conditions and visual appearance~\cite{NEURIPS2021_b0f2ad44, GU2023102904, zhang2024disentangling, MULLER2025103628}. Nervertheless, learning this decomposition is non-trivial because the model is not explicitly told which variations should be attributed to anatomy and which should be attributed to acquisition appearance. As illustrated in Fig.~\ref{teaser}(b), the same image pattern may admit multiple content and style explanations, making reliable factor assignment difficult and leading to a \textbf{factor identifiability} challenge~\cite{LIU2022102516, 10634512}. This challenge is especially pronounced in medical imaging, where style differences are often subtle, so that the same anatomy may appear differently across centers while its geometry remains unchanged. For instance, in contrast enhanced CT, hyperattenuation may arise either from true lesion enhancement as content or from contrast timing, kVp, or reconstruction kernel as style. Without explicit disentanglement, these competing explanations can easily become mixed, making the learned factors unreliable for both synthesis and analysis.

\IEEEpubidadjcol 
Motivated by the above analysis, we present MeDUET, a 3D \textbf{Me}dical image \textbf{D}isentangled \textbf{U}nifi\textbf{E}d Pre\textbf{T}raining model in the variational autoencoder (VAE) latent space, designed for both synthesis and analysis, as shown in Fig.~\ref{teaser}(c). This latent space enables compact spatial tokenization of 3D volumes while naturally interfacing with latent diffusion. \textit{Our key insight is that unified pretraining under style-shifted multi-center data should be treated as a factor identifiability problem, where content should capture anatomy and style should capture acquisition-related appearance.} Here, identifiability is used in an empirical sense, referring to whether the learned factors behave in a stable and selective manner under controlled diagnostics rather than whether they are uniquely recoverable in a formal theoretical sense. 

A further challenge is that, even when content and style are conceptually distinguishable, learning them in a stable and selective manner remains difficult in practice. (i) The model is not explicitly told which variations should be attributed to anatomy and which should be attributed to acquisition appearance, so factor assignment is inherently ambiguous. (ii) Once tokens from different samples are mixed, local evidence also becomes less reliable, increasing ambiguity and factor leakage. (iii) Moreover, even with improved local assignment, the learned factor spaces may still be poorly organized, causing content to remain sensitive to style changes and style to be insufficiently separated from anatomical variation. MeDUET addresses these challenges in a progressive manner. Concretely, to resolve factor assignment ambiguity, we introduce the demixing module that provides controlled mixtures to make factor supervision more explicit.  Mixed Factor Token Distillation (MFTD) is devised to reduce ambiguity in mixed regions by guiding each factor to remain faithful to its own source, thereby mitigating the 
mixed-region leakage issue. Ultimately, Swap-invariance Quadruplet Contrast (SiQC) further structures the factor spaces by promoting content invariance across style changes and style selectivity under content variation, thereby 
organizing the resulting factor spaces toward stable 
invariance and discriminability. With these more disentangled factors, MeDUET can be transferred directly to downstream synthesis and analysis. Our contributions can be summarized as follows: 
\begin{itemize}
\item We formulate unified pretraining for 3D medical image synthesis and analysis as a factor identifiability problem under multi-source data, and propose MeDUET, a VAE latent-space framework that learns domain-invariant content and domain-specific style for both downstream tasks.

\item We introduce an identifiability-driven SSL framework, where token demixing provides controllable supervision and is instantiated with an adversarial domain regularizer for basic content-style disentanglement. Building on this base factorization, MFTD reduces mixed-region ambiguity through source-faithful distillation, and SiQC promotes factor-wise invariance and discriminability.
% \item We introduce an identifiability-driven SSL framework, where demixing provides controllable supervision, MFTD reduces ambiguity in mixed regions through source-faithful distillation, and SiQC promotes factor-wise invariance and discriminability.

\item Comprehensive evaluation demonstrates that the identifiable factors learned by MeDUET transfer effectively to both synthesis and analysis across 6 datasets, 4 tasks, and 3 modalities, including \modCT{}, \modMRI{}, and \modPET{}, leading to improved generation quality, convergence speed, and controllability for synthesis, while offering competitive domain generalization and data efficiency for analysis.
\end{itemize}

The rest of the paper is organized as follows. Sec.~\ref{sec:related} reviews related work. Sec.~\ref{sec:method} presents our proposed MeDUET framework. Sec.~\ref{sec:exp} describes the experimental settings and reports comprehensive results, including comparisons with state-of-the-art (SOTA) methods, ablation studies, and in-depth mechanism analysis. Sec.~\ref{sec:discussion} discusses the implications and limitations. Finally, Sec.~\ref{sec:conclusion} concludes the paper.

\section{Related Work}
\label{sec:related}

\begin{figure*}
\centering
\includegraphics[width=1.0\linewidth]{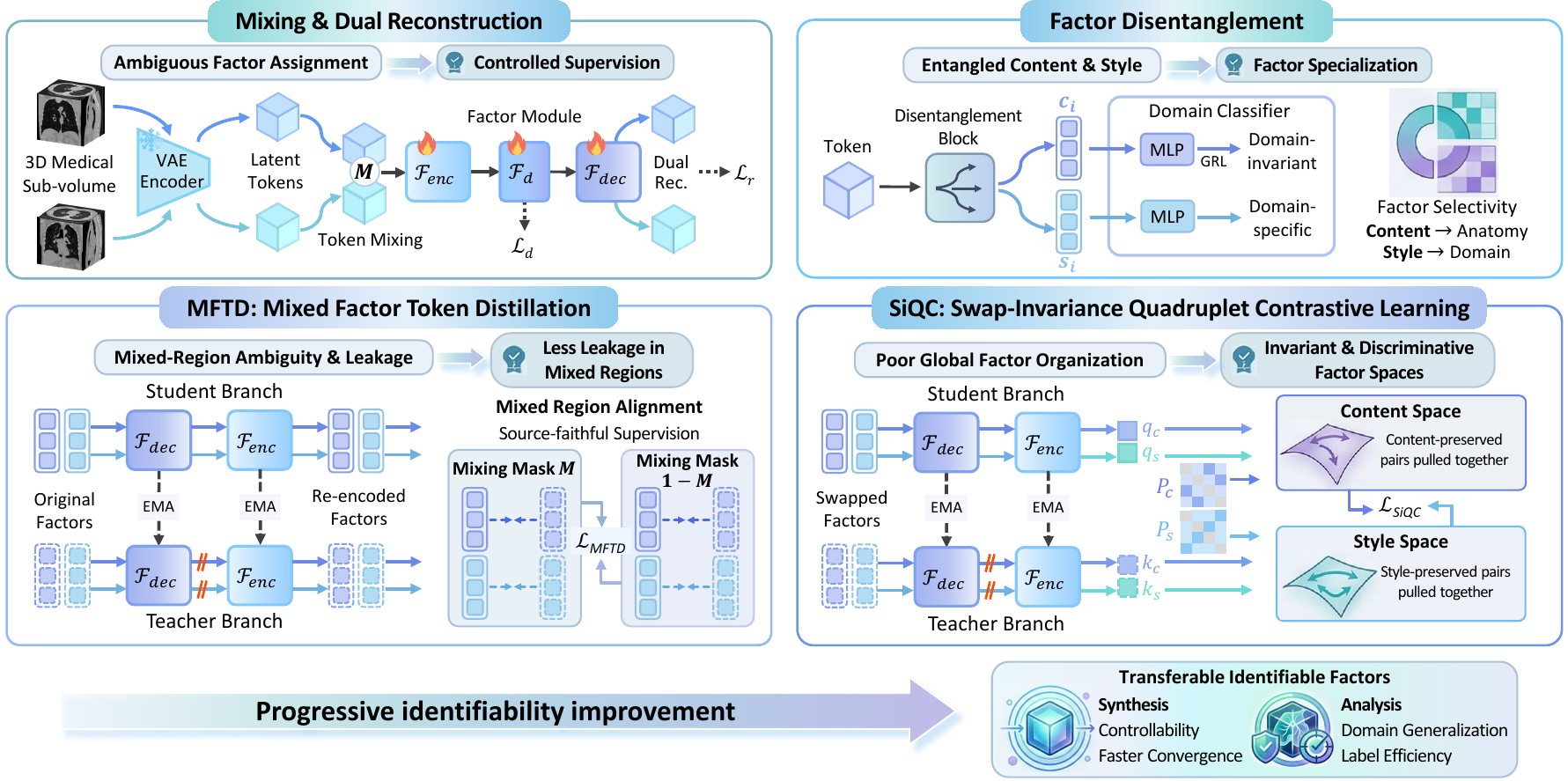}
\caption{\textbf{Overview of MeDUET.} \textbf{(a) Mixing \& Dual Reconstruction} provides controlled demixing supervision when factor assignment is unspecified. \textbf{(b) Factor Disentanglement} separates latent patches into anatomy-preserving content and acquisition-related style with factor specialization. \textbf{(c) MFTD} reduces mixed-region ambiguity and factor leakage through source-faithful supervision. \textbf{(d) SiQC} organizes the factor spaces with factor-wise invariance and discriminability under controlled swapping. These components jointly produce more selective and transferable factors.}
\label{framework}
% \vspace{-4pt}
\end{figure*} 

\subsection{3D Medical Image Synthesis} 
3D medical image synthesis aims to generate realistic volumetric data for clinical applications~\cite{11197215}. Benefiting from the development of diffusion models, recent studies have significantly advanced this field and demonstrated strong capabilities in modeling complex anatomical structures and fine-grained details~\cite{11311443, 10943915}. Beyond unconditional or conditional image generation, diffusion-based medical synthesis has also been explored in diverse clinical scenarios, including data augmentation~\cite{11175343, 11063450}, CT reconstruction~\cite{chung2024decomposed}, and counterfactual generation~\cite{Yeganeh_2025_CVPR}. These advances suggest that diffusion models provide a flexible and powerful paradigm for 3D medical image synthesis. In this work, we initialize diffusion models with pretrained SSL models in the VAE latent space, which not only accelerates convergence but also improves synthesis quality. Furthermore, the content and style factors learned by MeDUET provide a natural basis for more controllable medical image generation.

\subsection{3D Medical Image Analysis} 
Due to the cost of expert annotation, self-supervised pretraining on large-scale unlabeled 3D data has become an important paradigm for improving medical image analysis~\cite{11371598}. Existing methods are mainly built upon contrastive learning (CL)~\cite{He_2023_CVPR, Wu_2024_CVPR} or masked image modeling (MIM)~\cite{10977020, fu2026medgmae}. CL-based approaches typically learn invariant features through carefully designed positive and negative pairs, often relying on strong volumetric augmentations to encourage semantic consistency~\cite{11371598}. In contrast, MIM-based methods randomly mask or remove 3D patches and train models to reconstruct the missing content, thereby promoting contextual understanding of volumetric structures~\cite{TANG2026103770}. Despite their success, many existing methods still tend to encode scanner-dependent appearance variations together with anatomical semantics~\cite{TANG2026103770}, leading to entangled representations that may generalize poorly under domain shifts or limited-label settings. By comparison, our method explicitly disentangles anatomical content from imaging style, aiming to learn more robust and domain-consistent representations and thereby improving out-of-distribution (OOD) performance and data efficiency.

% \subsection{Diffusion Models for Representation Learning} 
% Owing to their strong generative capability across a wide range of visual synthesis tasks~\cite{hu2025lamd, bogensperger2025flowsdf, Peebles_2023_ICCV}, diffusion models have recently attracted growing interest as a foundation for representation learning~\cite{Wei_2023_ICCV, Hudson_2024_CVPR, Chu_2025_ICCV}. Emerging studies suggest that, beyond sample generation, diffusion models can encode semantically meaningful and transferable features, highlighting their potential as effective visual learners~\cite{mukhopadhyay2024text, 10938258}. In parallel, several works have further adapted diffusion architectures or training objectives to enhance their discriminative ability for visual perception and recognition tasks~\cite{Xiang_2023_ICCV, chen2025deconstructing, Ravishankar_2025_CVPR}. These advances indicate that diffusion models provide a promising bridge between generative modeling and visual understanding. However, their application to large-scale medical imaging, especially for jointly supporting generative and analytical tasks, remains largely underexplored.

\subsection{Disentangled Representation Learning} 
Disentangled representation learning aims to separate the underlying explanatory factors of observed data into interpretable and informative components~\cite{10579040, pmlr-v97-locatello19a}. Such representations are attractive for improving controllability, robustness, and transferability across downstream tasks. Prior studies have mainly focused on strengthening the disentanglement ability of generative models to isolate distinct variation factors from data distributions~\cite{pmlr-v80-kim18b, NEURIPS2024_96471570}. In medical imaging, disentangled learning has been explored for multiple applications~\cite{10634512}. Different from existing methods that are typically designed for a specific downstream task or standalone generation setting, MeDUET formulates disentanglement as a unified pretraining objective for both medical image synthesis and analysis. Moreover, MeDUET explicitly learns transferable content and style codes and verifies their utility through both generative conditioning and downstream analysis.

\section{Method}
\label{sec:method}
The framework of MeDUET is presented in Fig.~\ref{framework}. MeDUET follows a failure-mode-driven design principle to improve empirical factor identifiability under heterogeneous multi-source data. Token demixing (Sec.~\ref{sec3_1}) provides source-aware supervision for ambiguous factor assignment, while an adversarial domain regularizer in the factor module (Sec.~\ref{sec3_2}) encourages content-style separation. MFTD (Sec.~\ref{sec3_3}) reduces mixed-region ambiguity and factor leakage through source-faithful distillation, and SiQC (Sec.~\ref{sec3_4}) further organizes the factor spaces by promoting factor-wise invariance and discriminability under controlled swapping.
% The framework of MeDUET is presented in Fig.~\ref{framework}. MeDUET is guided by improving factor identifiability empirically under heterogeneous multi-source data. Token demixing (Sec.~\ref{sec3_1}) provides controllable supervision by explicitly specifying the source of content and style. MFTD (Sec.~\ref{sec3_3}) addresses mixed-region ambiguity by enforcing source-faithful factor assignment and reducing factor leakage, while SiQC (Sec.~\ref{sec3_4}) further promotes factor-wise invariance and discriminability under factor swapping. 
% Based on these disentangled and identifiable factors, downstream adaptation directly enables controllable synthesis and style-aware transfer.

% The overall framework of MeDUET is presented in Fig.~\ref{framework}, which is based on the VAE latent space, disentangling content from style and using token demixing as supervision. A student-teacher design supports two auxiliary proxy tasks, MFTD and SiQC, which provide targeted guidance to mixed tokens and enforce swap-invariant structure.

\subsection{Demixing for Identifiable Factor Supervision}
\label{sec3_1}
Under multi-center data with style variation, factor disentanglement is difficult because no explicit supervision specifies which variations to encode as content or style. Direct reconstruction or weak domain supervision alone therefore remains insufficient for factor assignment. To address this issue, we introduce a token demixing module that constructs and inverts controlled factor mixtures consistent with real clinical shifts, providing identifiable supervision for disentanglement. Specifically, given an input volume, we first obtain its latent representation via the frozen VAE encoder, and all subsequent operations are performed in latent space. Two sets of latent volume patches $\{\mathbf{z}_i, \mathbf{z}_j\}$ are sampled from two random volumes within a batch. Following the MIM paradigm, we then generate a mixed patch using a mixing function $\phi_m$ that combines the corresponding visible cubic tokens from $\mathbf{z}_i$ and $\mathbf{z}_j$ according to a binary mask $M$. The ViT encoder $\mathcal{F}_{enc}$ then takes the mixed tokens as input for representation learning. An unmixing function $\phi_u$ restores tokens to their original spatial positions based on the mixing notation $M$. The factor disentanglement module $\mathcal{F}_d$ is then applied, while the lightweight decoder $\mathcal{F}_{dec}$ reconstructs the original volumetric patches $\{\hat{\mathbf{z}}_i, \hat{\mathbf{z}}_j\}$ from the visible tokens. Finally, the dual reconstruction loss is formulated as
\begin{equation}
\mathcal{L}_r = \| (\hat{\mathbf{z}}_i - \mathbf{z}_i) \odot M \|_2^2 + \| (\hat{\mathbf{z}}_j - \mathbf{z}_j) \odot (1 - M) \|_2^2.  
\label{rec}
\end{equation}

\subsection{Factor Disentanglement}
\label{sec3_2}

The factor disentanglement module $\mathcal{F}_d$ provides a standard base factor-specialization step by decomposing the encoder output into two factors that serve complementary purposes: an anatomical content representation $\mathbf{c}_i \in \mathbb{R}^{L \times D_c}$ and an acquisition-related style representation $\mathbf{s}_i \in \mathbb{R}^{L \times D_s}$, where $L$ is the token length. We implement $\mathcal{F}_d$ with a lightweight convolutional layer. We further introduce domain classifiers for both factor learning branches. The content branch is trained adversarially through a gradient reversal layer (GRL)~\cite{JMLR:v17:15-239}, so that the encoder removes domain shift information from $\mathbf{c}_i$, while the style branch is trained normally to make $\mathbf{s}_i$ discriminative of the domain. Specifically, each branch uses a three-layer MLP domain classifier with a cross-entropy loss
\begin{align}
\resizebox{0.90\linewidth}{!}{
$
\mathcal{L}_d(i) = -\log\left([\sigma(\mathcal{G}_c(\mathcal{R}(\bar{\mathbf{c}}_i)))]_{y_i}\right) - \log\left([\sigma(\mathcal{G}_s(\bar{\mathbf{s}}_i))]_{y_i}\right),
$
}
\label{domain_cls}
\end{align}
where $y_i$ is the domain label, $\bar{\mathbf{c}}_i$ and $\bar{\mathbf{s}}_i$ denote the factors after token-level pooling, $\mathcal{G}_c$ and $\mathcal{G}_s$ are the domain classifiers for the content and style branches, respectively, and $\mathcal{R}$ represents the GRL. 
% This objective encourages the content branch to become domain-invariant, since the GRL drives it to confuse the domain classifier, while the style branch is encouraged to preserve domain-sensitive cues. 
We perform the same operation on $\mathbf{z}_j$, yielding $\mathcal{L}_d(j)$. The overall domain classifier loss is expressed as $\mathcal{L}_d = \mathcal{L}_d(i) + \mathcal{L}_d(j)$. Then the two factor tokens are integrated via another convolutional layer to obtain aggregated tokens for the subsequent reconstruction step. 

\subsection{MFTD for Source-faithful Factor Assignment}
\label{sec3_3}
Despite the fact that demixing provides supervision, the mixed regions no longer correspond to a single clean source, making local evidence ambiguous and factor assignment unreliable. Demixing supervision alone therefore remains insufficient to prevent factor leakage under incomplete context. To address this, we propose MFTD, which provides factor-specific teacher supervision, where content is distilled from the content source and style is distilled from the style source. In this way, MFTD encourages each factor to remain faithful to its own source even under incomplete contextual information, thereby strengthening disentanglement at the factor level.

Concretely, given the recovered patch tokens $\hat{\mathbf{z}}_i$ and $\hat{\mathbf{z}}_j$, we pass them through $\mathcal{F}_{enc}$ and $\mathcal{F}_d$ again without mixing, yielding factor tokens $\{\hat{\mathbf{c}}_i, \hat{\mathbf{s}}_i\}$ and $\{\hat{\mathbf{c}}_j, \hat{\mathbf{s}}_j\}$ that represent the full patch factors. We further introduce an Exponential Moving Average (EMA) teacher model, which performs the same re-encoding and re-disentanglement process to produce $\{\hat{\mathbf{c}}_i^{T}, \hat{\mathbf{s}}_i^{T}\}$ and $\{\hat{\mathbf{c}}_j^{T}, \hat{\mathbf{s}}_j^{T}\}$. These teacher factors serve as more stable global contextual priors for supervising the student factors. For content tokens, we compute the teacher-student discrepancy as
\begin{equation}
\Delta_i^c = \frac{1}{D_c} \| \hat{\mathbf{c}}_i - \mathrm{sg}[\hat{\mathbf{c}}_i^T] \|_2^2, \quad
\Delta_j^c = \frac{1}{D_c} \| \hat{\mathbf{c}}_j - \mathrm{sg}[\hat{\mathbf{c}}_j^T] \|_2^2,
\end{equation}
where $\mathrm{sg}[\cdot]$ denotes the stop-gradient operation. $\Delta_i^c$ and $\Delta_j^c$ measure how well the student content factors match the corresponding teacher factors from the two unmixed sources. 
% Averaging over the factor dimension mitigates the scale effect of high-dimensional representations and yields a token-wise discrepancy map.

We then apply the mixing mask $M$ to enforce source-consistent supervision in the mixed regions:
\begin{equation}
\mathcal{L}_{\mathrm{MFTD}}^c = 
\frac{\| \Delta_i^c \odot M + \Delta_j^c \odot (1 - M) \|_1}{L},
\end{equation}
where the region inherited from source $i$ is supervised by the teacher content of source $i$, while the complementary region is supervised by that of source $j$. Intuitively, this forces the student to preserve the correct content origin at each mixed location, instead of relying on ambiguous mixed context.

By combining it with the similarly computed style loss, the total token distillation loss is $\mathcal{L}_{\mathrm{MFTD}} = \lambda_c \mathcal{L}_{\mathrm{MFTD}}^c + \mathcal{L}_{\mathrm{MFTD}}^s$. To sum up, MFTD separately anchors content and style to their own teacher targets, which helps reduce cross-factor contamination and makes the disentangled factors more stable and better separated empirically.

\subsection{SiQC for Factor-wise Invariance and Discriminability}
\label{sec3_4}
Although demixing and MFTD improve local factor assignment, the factor spaces may still lack global organization. Specifically, content should remain invariant to style changes, while style should remain selective under content variation. To this end, we introduce SiQC, which constructs controlled swapped views and contrasts them in a factor-wise manner, thereby promoting factor-level invariance and discriminability.

% Although demixing and MFTD improve factor assignment, the resulting factor spaces may still lack clear global organization. Features with the same content can remain unstable across style variations, while features with the same style can remain insufficiently separated from content changes, which generic contrastive objectives fail to address explicitly. To address this issue, SiQC employs a swap-invariant quadruplet objective that pulls together pairs with the same content but different styles, and pairs with the same style but different contents, while pushing apart cross-factor negatives. This design explicitly structures both factor spaces to promote factor-level invariance and discriminability.

% \noindent\textbf{Content and style swapping.} 
Given the learned factors, we exchange factors between two input volumes, producing $\{\mathbf{c}_i, \mathbf{s}_j\}$ and $\{\mathbf{c}_j, \mathbf{s}_i\}$, which are then fed into $\mathcal{F}_{dec}$ to generate recovered patches $\mathbf{z}_{ij}$ and $\mathbf{z}_{ji}$, respectively. Next, $\mathcal{F}_{enc}$ and $\mathcal{F}_d$ are utilized again for re-encoding without mixing, yielding $\{\mathbf{c}_{ij}, \mathbf{s}_{ij}\}$ and $\{\mathbf{c}_{ji}, \mathbf{s}_{ji}\}$. The former retains the anatomical structure information of $\mathbf{z}_i$ and domain style characteristics of $\mathbf{z}_j$, while the latter does the opposite. In this way, the swapped views provide controlled pairs in which one factor is preserved and the other is intentionally changed. Let the two inputs be $\mathbf{z}_i$ and $\mathbf{z}_j$, and we form the quadruplet $\mathcal{V}(i,j) = \{\mathbf{z}_i, \mathbf{z}_j, \mathbf{z}_{ij}, \mathbf{z}_{ji}\}$. Taking content space as an example, $\mathbf{q}_c(a)$ and $\mathbf{k}_c(b)$ are defined as the pooled and normalized student and teacher content features for view $a, b \in \mathcal{V}(i,j)$. We define the binary positive mask as
\begin{equation}
\delta_c(\mathbf{z}_i) = \delta_c(\mathbf{z}_{ij}) = i, \quad 
\delta_c(\mathbf{z}_j) = \delta_c(\mathbf{z}_{ji}) = j,
\end{equation}
\begin{equation}
P_c(a, b) = \mathbf{1}\{\delta_c(a) = \delta_c(b)\} \cdot \mathbf{1}\{a \ne b\},
\end{equation}
where $\delta_c(\cdot)$ is the content identity map, and $\mathbf{1}$ is the indicator function. This means that each original view and its content-preserved swapped counterpart are treated as positives, since they share the same underlying content despite differing in style. Then the similarity and loss are defined as
\begin{align}
\resizebox{0.90\linewidth}{!}{
$
S_c(a, b) = \exp(\alpha_c)\, \mathbf{q}_c(a) \cdot \mathrm{sg}[\mathbf{k}_c^T(b)], \quad S_c(a, a) = -\infty,
$
}
\end{align}
\begin{equation}
\ell_c(a, b) = S_c(a, b) - \log \sum_{b' \in \mathcal{V}(i,j)\setminus\{a\}} \exp(S_c(a, b')),    
\end{equation}
$S_c(a,b)$ measures the similarity of two views, and $\ell_c(a,b)$ encourages a view to stay close to its content-matched positives while being separated from the remaining views.

\begin{equation}
\mathcal{L}_{\mathrm{SiQC}}^c 
= \mathbb{E}_{a \in \mathcal{V}(i,j)} 
\!\left[
-\frac{\sum_{b \in \mathcal{V}(i,j)} P_c(a,b)\, \ell_c(a,b)}
{\sum_{b \in \mathcal{V}(i,j)} P_c(a,b)}
\right],
\end{equation}
This objective pulls $\{\mathbf{z}_i, \mathbf{z}_{ij}\}$ and $\{\mathbf{z}_j, \mathbf{z}_{ji}\}$ together in content space, since each pair shares the same content but differs in style. Then the total loss of SiQC is formulated as $\mathcal{L}_{\mathrm{SiQC}} 
= \mathcal{L}_{\mathrm{SiQC}}^c + \mathcal{L}_{\mathrm{SiQC}}^s$, where $\alpha_c$ is a learnable scale. The full objective applies the same idea symmetrically to both content and style, so that each factor space is encouraged to preserve its own semantics while remaining discriminative to the other factor. During SiQC, teacher features are stop-gradient, while student features and the student path that produces swapped views remain fully differentiable, so that SiQC improves both factor encoders and the swap mechanism.

The total loss of MeDUET pretraining is formulated as 
\begin{equation}
\mathcal{L} = \mathcal{L}_r + \lambda_1 \mathcal{L}_d + \lambda_2 \mathcal{L}_{\mathrm{MFTD}} + \lambda_3 \mathcal{L}_{\mathrm{SiQC}}.
\label{total_loss}
\end{equation}
% where $\lambda_i$ are coefficients to balance loss contribution. 
where the four terms address complementary factor-identifiability failures: source-aware demixing reconstruction, directional factor separation, mixed-region leakage, and global factor-space organization, respectively. Table~\ref{tab:failure_mode_design} summarizes these objective-level roles, where the main proposed identifiability mechanisms are demixing, MFTD, and SiQC.

\begin{table}[t]
\centering
\setlength{\tabcolsep}{1.5mm}
\renewcommand{\arraystretch}{1.0}
\caption{\textbf{Failure-mode-driven design of identifiability mechanisms.} Each term in the pretraining objective targets a distinct empirical factor-identifiability failure mode.}
\resizebox{1.0\linewidth}{!}{
\begin{tabular}{p{0.24\linewidth} p{0.38\linewidth} p{0.22\linewidth} p{0.16\linewidth}}
\toprule
\textbf{Failure mode} & \textbf{Remedy (Mechanism)} & \textbf{Extra trainable params} & \textbf{Evidence} \\
\midrule
No supervision specifying content vs. style &
Token demixing provides source-aware supervision (Sec.~\ref{sec3_1}, $\mathcal{L}_r$). &
None &
Tables~\ref{ablation}, \ref{linear} \\
\midrule
Content and style are not directionally separated &
Base adversarial factor disentanglement (Sec.\ref{sec3_2}, $\mathcal{L}_d$) &
Factor conv and domain heads, pretraining-only &
Tables~\ref{ablation}, \ref{linear} \\
\midrule
Mixed-region ambiguity and factor leakage &
MFTD anchors each factor to its source-faithful teacher target (Sec.~\ref{sec3_3}, $\mathcal{L}_\mathrm{MFTD}$). &
None, non-trainable EMA teacher &
Tables~\ref{ablation}, \ref{linear}, \ref{linear_OOD}; Fig.~\ref{factor_matrix} \\
\midrule
Poorly organized global factor spaces &
SiQC enforces factor-wise invariance and selectivity (Sec.~\ref{sec3_4}, $\mathcal{L}_\mathrm{SiQC}$). &
Learnable scales only &
Tables~\ref{ablation}, \ref{linear}, \ref{linear_OOD}; Fig.~\ref{factor_matrix} \\
\bottomrule
\end{tabular}}
\label{tab:failure_mode_design}
\vspace{-24pt}
\end{table}

\subsection{Downstream Readouts of Identifiable Factors}
\label{sec3_5}
After pretraining, we adapt the pretrained backbone to downstream architectures, derive volume-level factors from the frozen model, and exploit them in task-specific ways. In synthesis, the factors provide controllable conditional signals. In analysis, they enable style-aware augmentation.

\subsubsection{Transferring to Downstream Models}
For generative tasks, we modify the pretrained ViT model to adapt and initialize the DiT/SiT~\cite{Peebles_2023_ICCV, ma2024sit} models for diffusion-based generation. Following~\cite{Chu_2025_ICCV}, we reintroduce the shift ($\beta$) and scale ($\gamma$) parameters to implement AdaLN-Zero and layer normalization blocks as conditional input modules. For analysis tasks, we adopt modified UNETR~\cite{Hatamizadeh_2022_WACV} as the backbone, which seamlessly inherits the pretrained ViT weights from MeDUET.

\subsubsection{Volume-level Factor Generation}
We extract the content and style vectors by cropping each volume into fixed-size sub-volumes and feeding into the frozen MeDUET model, where the factor disentanglement module produces content and style representations $\mathbf{c} \in \mathbb{R}^{L \times D_c}$ and $\mathbf{s} \in \mathbb{R}^{L \times D_s}$. The resulting content and style representations from all sub-volumes are then aggregated at the volume level. With the sub-volumes covering the full volume, we apply patch-level and token-level mean pooling to the aggregated representations, yielding the final content and style vectors $\mathbf{c}_0 \in \mathbb{R}^{D_c}$ and $\mathbf{s}_0 \in \mathbb{R}^{D_s}$.

\subsubsection{Controllable Synthesis with Dual Conditional Diffusion}
For synthesis, we use the volume-level content and style vectors as two independent conditional signals for DiT/SiT, which are injected through a dual-branch AdaLN-Zero module to separately modulate anatomical content and acquisition appearance. At inference, dual classifier-free guidance (CFG)~\cite{ho2022classifier} is applied to independently adjust content and style strengths, enabling controllable generation with disentangled anatomical and appearance control. Detailed information is provided in the supplementary material.

\subsubsection{Style-Aware Analysis via Latent Augmentation}
For analysis, we exploit the learned style factors to perform latent style augmentation by preserving source content while shifting its style toward a target-domain style prototype estimated from unlabeled target volumes. The original and style-augmented latents are supervised with the same labels, encouraging the downstream model to focus on anatomical content and improving robustness to style/domain shifts. Please refer to the supplementary material for details.

\section{Experiments}
\label{sec:exp}
\subsection{Experimental Setup}
\subsubsection{Datasets} 
We employ the VoCo-10k dataset~\cite{Wu_2024_CVPR} for pretraining, which includes 10 public CT datasets covering diverse sources and anatomical regions~\cite{landman2015miccai, 8458220, simpson2019large, clark2013cancer, SETIO20171, doi:10.1148/radiol.2021210384, ma2024automatic, https://doi.org/10.1118/1.3528204, grossberg2018imaging, wasserthal2023totalsegmentator}, consisting of 10,500 CT scans in total. We follow the previously released open-source data splits~\cite{Wu_2024_CVPR, 11004165, 10977020} for the pretraining datasets. For downstream experiments, all diffusion models are trained using the VoCo-10k dataset. Additionally, we evaluate MeDUET on multiple analysis datasets spanning \modCT{}, \modMRI{}, and \modPET{} modalities, including BTCV~\cite{landman2015miccai}, WORD~\cite{LUO2022102642}, AMOS~\cite{NEURIPS2022_ee604e1b}, BraTS 21~\cite{baid2021rsna} and AutoPET III~\cite{dexl2026autopet3} for segmentation tasks, and CC-CCII~\cite{zhang2020clinically} for the COVID-19 classification task. We also follow the data splits used in prior work~\cite{Wu_2024_CVPR, TANG2026103770} to ensure fair comparison. Details of the datasets are provided in the supplementary material.

\subsubsection{Implementation Details} 
We employ MAISI-VAE~\cite{10943915} as the tokenizer to generate volume latents, which are cached to facilitate efficient training and inference. The pretraining process consists of 200k steps, using the AdamW optimizer with a cosine learning rate schedule.
% The isotropic voxel spacing is set to 1.5~mm, and each subvolume patch has a size of $96 \times 96 \times 96$. 
We introduce the MFTD module at 40k steps when the teacher model has become sufficiently stable. We treat each dataset as a domain because it typically reflects a distinct center/protocol/scanner and thus serves as a stable proxy for style shifts. All experiments are conducted on four NVIDIA A100 GPUs. We adopt ViT-B as the backbone encoder of MeDUET. The dimensions of the content and style representations, $D_c$, $D_s$, are set to 768 and 192, respectively. The hyperparameters of the loss functions are set as follows: $\lambda_1=0.2, \lambda_2=0.5, \lambda_3=0.3$, and $\lambda_c=0.5$. 
% The GRL strength of the domain classifier is set to 1.0. For the EMA teacher, the decay rate is initialized at 0.997 and gradually increased to 0.9997 during pretraining using a cosine schedule. 

% \begin{table}[t]
% \centering
% \caption{The overview of pre-processing and pretraining settings in the experiments.}
% \setlength{\tabcolsep}{1.0mm}
% % \renewcommand{\arraystretch}{1.15}
% % \resizebox{0.9\linewidth}{!}{
% \begin{tabular}{lc}
% \toprule
% \multicolumn{2}{l}{\textbf{Pre-processing Settings}} \\
% \midrule
% Spacing          & $1.5\times1.5\times1.5$ \\
% Intensity        & $[-175, 250]$ \\
% Norm             & $[0, 1]$ \\
% Sub-volume size  & $96\times96\times96$ \\
% Latent size      & $4\times24\times24\times24$ \\
% Augmentation     & Random Rotate and Flip \\
% \midrule
% \multicolumn{2}{l}{\textbf{Pretraining Settings}} \\
% \midrule
% Pretraining steps   & 200k \\
% Optimizer           & AdamW \\
% Weight decay        & 1e-2 \\
% Optimizer momentum  & $\beta_1, \beta_2 = 0.9, 0.95$ \\
% Optimizer lr        & 1e-4 \\
% Batch size          & $64\times4 = 256$ \\
% Lr schedule         & Warmup cosine \\
% Warmup steps        & 2k \\
% Factor dimension    & $D_c, D_s = 768, 192$ \\
% Loss coefficients   & $\lambda_1, \lambda_2, \lambda_3, \lambda_c = 0.2, 0.5, 0.3, 0.5$ \\
% EMA decay           & 0.997, 0.9997 \\
% GRL coefficient     & 1.0 \\
% \bottomrule
% \end{tabular}
% % }
% \label{pretrain_settings}
% \end{table}

For synthesis, we condition DiT~\cite{Peebles_2023_ICCV} / SiT~\cite{ma2024sit} on the content and style, replacing coarse metadata with fine-grained control. We build our model upon the original implementations and adopt the same hyperparameter configurations. For analysis, following~\cite{Wu_2024_CVPR, 11004165, 10977020, TANG2026103770}, the fine-tuning setups are kept largely consistent with the pretraining configurations. For segmentation tasks, we modify UNETR~\cite{Hatamizadeh_2022_WACV} to perform standard multi-class segmentation directly in the original label space. Specifically, instead of passing the latent-space outputs through the VAE decoder to reconstruct binary masks, we replace the original prediction branch with a simple multi-class segmentation decoder that outputs voxel-wise logits. For classification tasks, we remove the VAE decoder and modify the final output layer of UNETR to directly produce the predictive logits. Details of the pre-processing and experimental settings are provided in the supplementary material.

\begin{table}[t]
\centering
\caption{Synthesis performance comparison. The best results are \textbf{bolded}, and the second best results are \underline{underlined}. $\dagger$: using pre-defined metadata vectors. $\ddagger$: using learned content and style vectors. 100k/200k: pretrained steps of MeDUET. We evaluate SiT B/4 + MeDUET for subsequent experiments, using pretraining weights from 200k steps by default.}
\setlength{\tabcolsep}{2.0mm}
\resizebox{1.0\linewidth}{!}{
\begin{tabular}{lccc}
\toprule
\textbf{Method} & \textbf{FID}\ $\downarrow$ & \textbf{MMD}\ $\downarrow$ & \textbf{MS-SSIM}\ $\downarrow$ \\
\midrule
\rowcolor{tablegray}
\multicolumn{4}{c}{\textbf{\textit{Medical Image Synthesis Models}}} \\
WDM~\cite{friedrich2024wdm} & 0.9668 & 0.6612 & 0.2284 \\
MedSyn~\cite{10566053} & 0.9873 & 0.6734 & 0.2325 \\
MAISI~\cite{10943915} & 0.9139 & 0.6292 & 0.2057 \\
3D MedDiffusion~\cite{11063450} & 0.9216 & 0.6327 & 0.2032 \\
\midrule
\rowcolor{tablegray}
\multicolumn{4}{c}{\textbf{\textit{General Diffusion Models}}} \\
DiT-B/4$^{\dagger}$  & 0.9207 & 0.6329 & 0.1957 \\
DiT-B/4$^{\ddagger}$ & 0.9074 & 0.6175 & 0.1906 \\
\rowcolor{tablelightblue}
+ MeDUET 100k/200k$^{\dagger}$  & 0.8763 / 0.8727 & 0.6097 / 0.6074 & 0.1892 / 0.1876 \\
\rowcolor{tableblue}
+ MeDUET 100k/200k$^{\ddagger}$ & 0.8642 / 0.8611 & 0.6028 / 0.6003 & 0.1824 / 0.1803 \\
\midrule
SiT-B/4$^{\dagger}$  & 0.8670 & 0.6023 & 0.1834 \\
SiT-B/4$^{\ddagger}$ & 0.8533 & 0.6012 & 0.1798 \\
\rowcolor{tablelightblue}
+ MeDUET 100k/200k$^{\dagger}$  & 0.8039 / 0.8011 & 0.5806 / 0.5782 & 0.1712 / 0.1692 \\
\rowcolor{tableblue}
\textbf{+ MeDUET 100k/200k}$^{\ddagger}$ & \underline{0.7908} / \textbf{0.7874} & \underline{0.5627} / \textbf{0.5598} & \underline{0.1677} / \textbf{0.1659} \\
\bottomrule
\end{tabular}
}
\label{synthesis}
% \vspace{-4pt}
\end{table}

\begin{figure}[t]
  \centering
  \includegraphics[width=\linewidth]{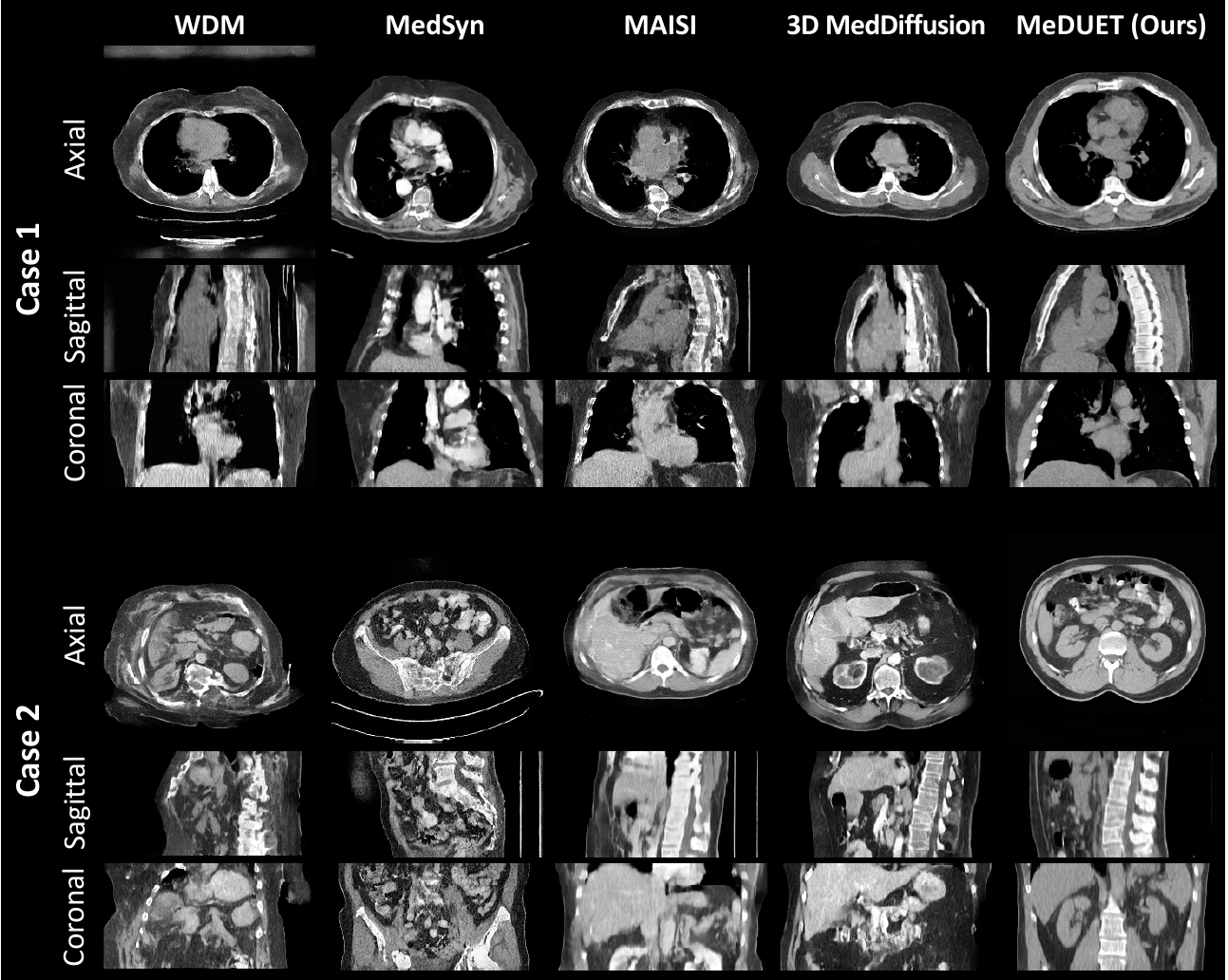}
  \caption{Qualitative comparison of synthesized volumes.}
  \label{visual}
  \vspace{-12pt}
\end{figure}

\begin{figure}[t]
    \centering
    \includegraphics[width=1.0\linewidth]{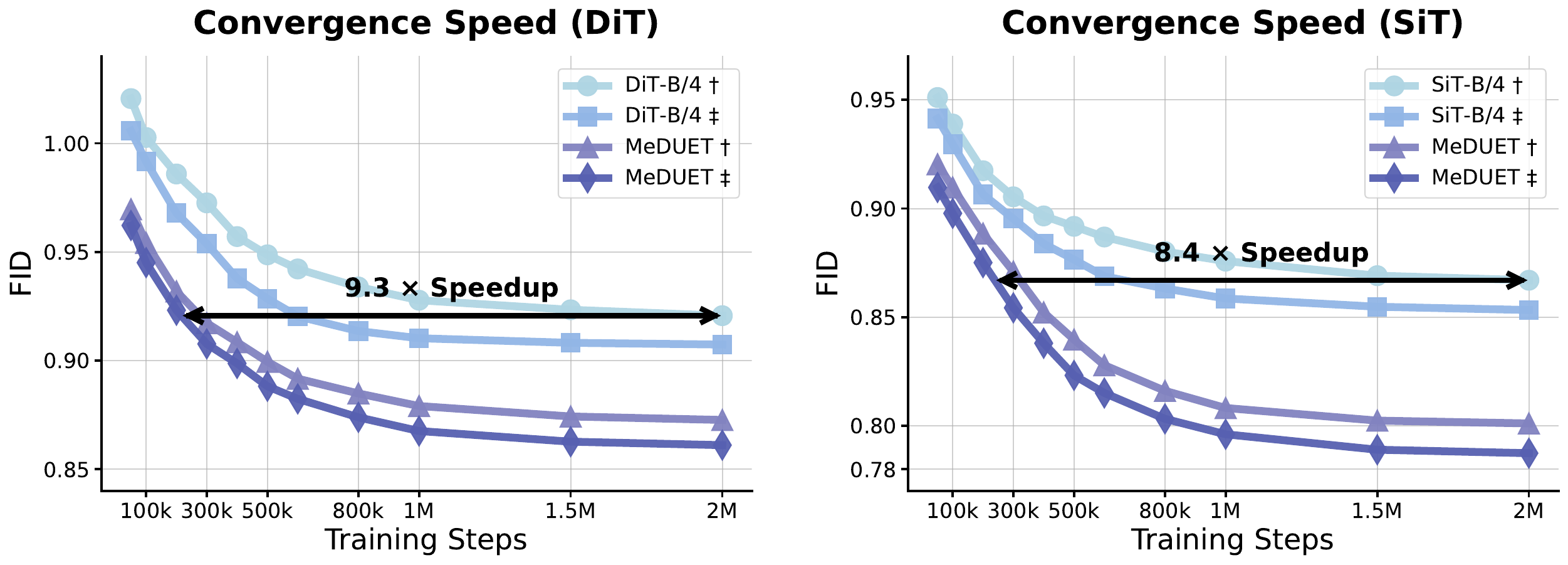}
    \caption{\textbf{Convergence speed comparison.} $\dagger$: using pre-defined metadata. $\ddagger$: using learned content and style.}
    \label{convergence}
    % \vspace{-4pt}
\end{figure}

\begin{table}[t]
\centering
\caption{Synthesis performance comparison for MRI synthesis. $\dagger$: using modality one-hot vectors. $\ddagger$: using content and style vectors.}
\setlength{\tabcolsep}{1.5mm}
\resizebox{0.90\linewidth}{!}{
\begin{tabular}{lcccc}
\toprule
\multirow{2}{*}{\textbf{Method}} 
& \multicolumn{3}{c}{\textbf{Image Quality Assessment}} 
& \multicolumn{1}{c}{\textbf{Synthetic-to-real Applicability}} \\
\cmidrule(lr){2-4} \cmidrule(lr){5-5}
& \textbf{FID}\ $\downarrow$ 
& \textbf{MMD}\ $\downarrow$ 
& \textbf{MS-SSIM}\ $\downarrow$ 
& \textbf{Dice Score (\%)}\ $\uparrow$ \\
\midrule
DiT-B/4$^{\dagger}$  & 2.5686 & 1.6514 & 0.2917 & 70.51\stdgray{7.1} \\
DiT-B/4$^{\ddagger}$ & 2.4753 & 1.6126 & 0.2870 & 72.28\stdgray{5.1} \\
\rowcolor{tablelightblue}
+ MeDUET$^{\dagger}$  & 2.4093 & 1.5307 & 0.2676 & 74.12\stdblue{5.0} \\
\rowcolor{tableblue}
+ MeDUET$^{\ddagger}$ & 2.3734 & 1.4863 & 0.2690 & 74.94\stdpurple{6.2} \\
\midrule
SiT-B/4$^{\dagger}$  & 2.4351 & 1.5072 & 0.2646 & 76.07\stdgray{5.8} \\
SiT-B/4$^{\ddagger}$ & 2.3526 & 1.4783 & 0.2539 & 76.22\stdgray{7.7} \\
\rowcolor{tablelightblue}
+ MeDUET$^{\dagger}$  & \underline{2.2029} & \underline{1.3796} & \underline{0.2518} & \underline{77.45}\stdblue{4.6} \\
\rowcolor{tableblue}
\textbf{+ MeDUET}$^{\ddagger}$ & \textbf{2.1760} & \textbf{1.3436} & \textbf{0.2423} & \textbf{78.78\stdpurple{5.5}} \\
\bottomrule
\end{tabular}
}
\label{synthesis_MRI}
% \vspace{-4pt}
\end{table}

\subsection{Medical Image Synthesis}
\label{sec4_2}
We integrate pretrained MeDUET into DiT~\cite{Peebles_2023_ICCV} and SiT~\cite{ma2024sit}, following their default settings. We reimplemented all baseline methods from scratch using the pretraining dataset. Notably, we leverage the learned content and style vectors as conditioning signals in CFG instead of pre-defined metadata vectors (e.g. body region, voxel spacing). Evaluations are conducted on the generated 1k volumes. For evaluation metrics, Fréchet Inception Distance (FID)~\cite{NIPS2017_8a1d6947} and Maximum Mean Discrepancy (MMD), which both utilize the MedicalNet 3D~\cite{chen2019med3d} as the feature extractor, are used to assess the fidelity and realism, while Multi-Scale Structural Similarity Index (MS-SSIM) is employed to evaluate the diversity. 

\subsubsection{Synthesis Quality Comparison}
As reported in Table~\ref{synthesis}, MeDUET achieves the best overall generative performance among the compared methods. These results indicate that MeDUET not only improves synthesis fidelity and realism, but also preserves stronger diversity in the generated volumes. Notably, replacing the previously used metadata vectors with our learned disentangled factors consistently leads to better performance, suggesting the informative conditioning of our factors. This advantage is especially important in medical imaging, where anatomical structure and acquisition style are often entangled and cannot be fully characterized by simple discrete labels. Fig.~\ref{visual} further provides qualitative comparisons, showing that our method generates more realistic structures and visually more convincing details than the baselines. 

\begin{figure*}[t]
  \centering
  \includegraphics[width=0.90\linewidth]{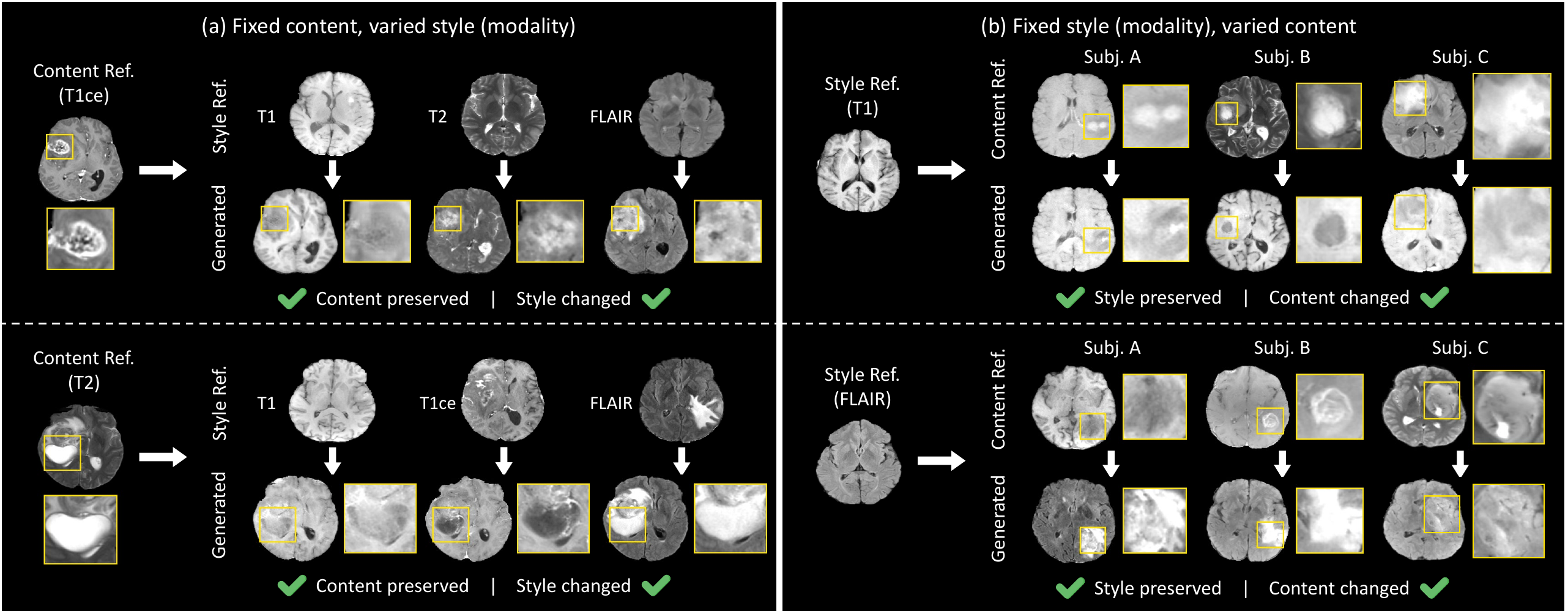}
  % \vspace{-4pt}
\caption{\textbf{Qualitative factor swapping examples for MRI synthesis.} (a) Fixed-content, varied-style generation. Given a content reference, MeDUET changes the target MRI modality while preserving subject-specific anatomy and tumor layout.  (b) Fixed-style, varied-content generation. Given a fixed modality/style reference, MeDUET generates different subject anatomies while maintaining the target modality appearance.}
  \label{vis_control}
\end{figure*}

\begin{figure}[t]
  \centering
  \includegraphics[width=1.0\linewidth]{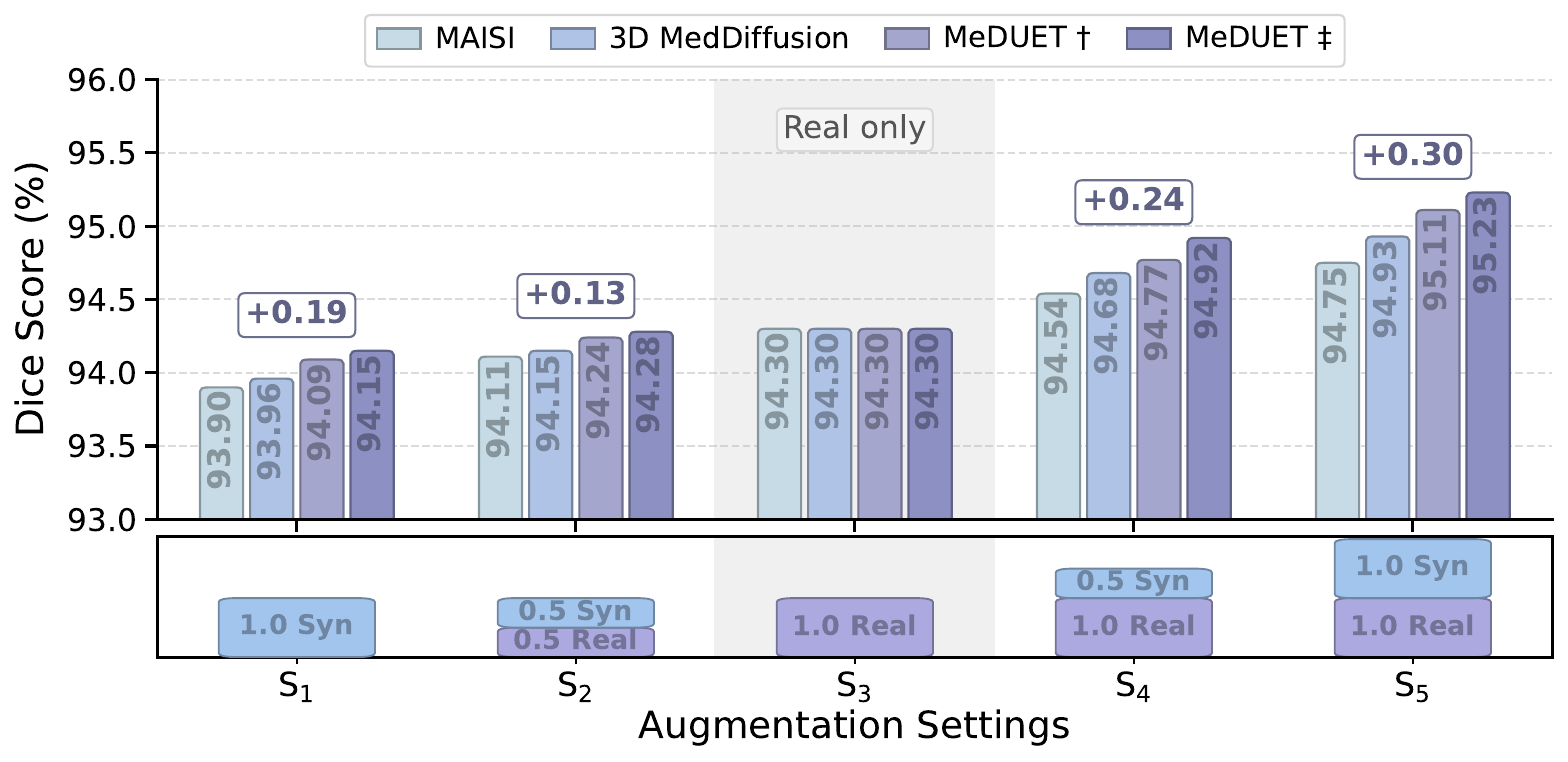}
  % \vspace{-4pt}
  \caption{Performance comparison under different data augmentation regimes. $\dagger$ uses pre-defined metadata vectors, and $\ddagger$ uses learned content and style vectors. S1 to S5 denote the real and synthetic data compositions.}
  \label{augmentation}
  % \vspace{-4pt}
\end{figure}

\subsubsection{Convergence Acceleration}
Fig.~\ref{convergence} reports the training dynamics of DiT, SiT, and their MeDUET initialized counterparts. Compared with training from scratch, MeDUET initialization substantially accelerates optimization, yielding a \textbf{9.3$\times$} faster FID convergence speed for DiT and an \textbf{8.4$\times$} speedup for SiT. This suggests that MeDUET provides a more favorable initialization for diffusion-based synthesis. The advantage becomes more pronounced when replacing pre-defined metadata with the disentangled factors and when extending the pretraining duration, further highlighting the benefit of our scheme and initialization strategy.

\subsubsection{Cross-Modality MRI Synthesis}
To evaluate transfer beyond the \modCT{} pretraining distribution, we conduct \modMRI{} synthesis on the combined BraTS 21~\cite{baid2021rsna} and BraTS 24~\cite{de20242024} datasets, using T1, T1ce, T2, and FLAIR as modality/style conditions. We compare modality one-hot conditioning, which assigns each modality a discrete vector, with MeDUET-extracted factors as learned conditions. We further report a synthetic-to-real Dice score, obtained by training nnU-Net~\cite{isensee2018nnu} on synthetic MRI volumes and testing it on held-out real BraTS volumes, to assess downstream segmentation utility. Table~\ref{synthesis_MRI} shows that MeDUET improves both DiT and SiT, and learned factors provide more effective guidance than one-hot conditions.

Fig.~\ref{vis_control} further provides qualitative factor-swapping examples to assess the controllability of MeDUET in MRI synthesis. When the content factor is fixed, MeDUET preserves subject-specific anatomy and tumor layout while changing the MRI modality appearance; conversely, fixing the style factor maintains modality-specific contrast across different subjects, supporting disentangled control over content and style.

\subsubsection{Conditional Generation for Data Augmentation}
Following~\cite{10943915, 11063450}, we integrate ControlNet~\cite{Zhang_2023_ICCV, Tan_2025_ICCV} using segmentation masks to improve controllability. Subsequently, the synthesized volumes are employed as augmented data to enhance segmentation performance. We train nnU-Net~\cite{isensee2018nnu} on TotalSegmentator~\cite{wasserthal2023totalsegmentator} under five training regimes. As illustrated in Fig.~\ref{augmentation}, our MeDUET consistently surpasses baselines across the five augmentation settings. Disentangled factors further boost segmentation, indicating anatomy-preserving content control and style-aware diversification, and validating our disentangled, controllable generation framework.

\begin{table*}[t]
\centering
\setlength{\tabcolsep}{1.5mm}
\caption{Results on the segmentation task under different data proportions. The fourth row denotes the data ratios for training. We report the mean Dice score (\%). The best results are \textbf{bolded}, and the second best results are \underline{underlined}. $\dagger$: without style augmentation. $\ddagger$: using style augmentation with learned style codes.}
\resizebox{1.0\linewidth}{!}{
\begin{tabular}{lccccc|ccccc|ccccc|ccccc|ccccc}
\toprule
\multirow{4}{*}{\textbf{Method}}
& \multicolumn{25}{c}{\textbf{Dice Score (\%) $\uparrow$}} \\
\cmidrule(lr){2-26}
\cmidrule(lr){2-26}

& \multicolumn{15}{c|}{\cellcolor{ctbg}{\color{cttxt}\textbf{CT}}}
& \multicolumn{5}{c|}{\cellcolor{mribg}{\color{mritxt}\textbf{MRI}}}
& \multicolumn{5}{c}{\cellcolor{petbg}{\color{pettxt}\textbf{PET}}} \\
\cmidrule(lr){2-16}\cmidrule(lr){17-21}\cmidrule(lr){22-26}

& \multicolumn{5}{c|}{\cellcolor{ctbg}{\color{cttxt}\textbf{BTCV}}}
& \multicolumn{5}{c|}{\cellcolor{ctbg}{\color{cttxt}\textbf{AMOS}}}
& \multicolumn{5}{c|}{\cellcolor{ctbg}{\color{cttxt}\textbf{WORD}}}
& \multicolumn{5}{c|}{\cellcolor{mribg}{\color{mritxt}\textbf{BraTS 21}}}
& \multicolumn{5}{c}{\cellcolor{petbg}{\color{pettxt}\textbf{AutoPET III}}} \\
\cmidrule(lr){2-6}\cmidrule(lr){7-11}\cmidrule(lr){12-16}\cmidrule(lr){17-21}\cmidrule(lr){22-26}
& 1-shot & 10\% & 50\% & 100\% & Avg
& 1-shot & 10\% & 50\% & 100\% & Avg
& 1-shot & 10\% & 50\% & 100\% & Avg
& 1-shot & 10\% & 50\% & 100\% & Avg 
& 1-shot & 10\% & 50\% & 100\% & Avg \\
\midrule
\rowcolor{tablegray}
\multicolumn{26}{c}{\textbf{\textit{From Scratch}}} \\
UNETR~\cite{Hatamizadeh_2022_WACV}
& 24.27 & 42.85 & 77.74 & 79.82 & 56.17
& 10.06 & 60.06 & 74.96 & 77.02 & 55.52
& 30.89 & 73.76 & 76.43 & 78.73 & 64.95
& 53.60 & 72.36 & 87.12 & 88.78 & 75.47
& 13.09 & 31.82 & 45.10 & 47.51 & 34.38 \\
Swin-UNETR~\cite{hatamizadeh2021swin}
& 27.71 & 51.33 & 78.83 & 80.53 & 59.60
& 9.59 & 63.45 & 80.16 & 82.51 & 58.93
& 43.68 & 74.03 & 78.71 & 80.39 & 69.20
& 55.95 & 72.66 & 87.39 & 89.01 & 76.25
& 15.83 & 30.49 & 44.83 & 46.83 & 34.50 \\
\midrule
\rowcolor{tablegray}
\multicolumn{26}{c}{\textbf{\textit{General SSL}}} \\
MAE3D~\cite{Chen_2023_WACV}
& 31.86 & 55.79 & 79.08 & 81.33 & 62.02
& 25.46 & 62.83 & 79.86 & 82.26 & 62.60
& 39.23 & 72.53 & 76.82 & 78.97 & 66.89
& 48.53 & 71.59 & 86.42 & 88.15 & 73.67
& 18.10 & 29.93 & 43.09 & 45.50 & 34.16 \\
SimMIM~\cite{Xie_2022_CVPR}
& 34.73 & 56.34 & 74.86 & 76.03 & 60.49
& 22.63 & 63.70 & 78.56 & 80.83 & 61.43
& 37.56 & 73.25 & 78.23 & 79.60 & 67.16
& 42.29 & 71.74 & 87.05 & 88.67 & 72.44
& 14.29 & 28.71 & 44.29 & 45.82 & 33.28 \\
MoCo v3~\cite{He_2020_CVPR}
& 26.82 & 38.27 & 77.91 & 79.54 & 55.64
& 19.67 & 65.76 & 79.66 & 81.95 & 61.76
& 41.43 & 72.89 & 79.19 & 80.73 & 68.56
& 36.72 & 72.16 & 86.72 & 88.32 & 70.98
& 12.48 & 29.56 & 43.37 & 46.23 & 32.91\\
\midrule
\rowcolor{tablegray}
\multicolumn{26}{c}{\textbf{\textit{Medical Image SSL}}} \\
Swin-UNETR~\cite{hatamizadeh2021swin}
& 28.52 & 48.31 & 78.59 & 81.54 & 59.24
& 18.43 & 65.49 & 79.64 & 80.97 & 61.13
& 48.16 & 75.07 & 80.37 & 81.79 & 71.35
& 47.17 & 70.53 & 87.15 & 88.94 & 73.45
& 17.10 & 31.59 & 45.06 & 48.50 & 35.56 \\
SwinMM~\cite{wang2023swinmm}
& 28.77 & 47.93 & 79.26 & 81.81 & 59.44
& 16.93 & 68.57 & 80.46 & 81.13 & 61.77
& 42.86 & 73.98 & 81.06 & 82.49 & 70.10
& 38.51 & 70.17 & 87.44 & 89.27 & 71.35
& 20.17 & 30.46 & 44.67 & 47.63 & 35.73 \\
GVSL~\cite{He_2023_CVPR}
& 24.86 & 41.79 & 78.93 & 81.87 & 56.86
& 10.84 & 63.45 & 80.28 & 81.38 & 58.99
& 37.87 & 74.10 & 80.78 & 82.36 & 68.78
& 60.20 & 74.51 & 85.94 & 87.62 & 77.07
& 23.89 & 34.73 & 44.16 & 47.35 & 37.53 \\
VoCo~\cite{Wu_2024_CVPR}
& 63.33 & 77.85 & 81.52 & 83.85 & 76.64
& 38.80 & 73.34 & 81.80 & 84.44 & 69.59
& 64.24 & 77.28 & 82.48 & 84.06 & 77.02
& 49.76 & 71.80 & 86.93 & 88.81 & 74.33
& 27.09 & 36.39 & 46.22 & 49.37 & 39.77 \\
GL-MAE~\cite{11004165}
& 52.46 & 65.83 & 80.14 & 82.01 & 70.11
& 38.17 & 72.43 & 81.96 & 84.20 & 69.19
& 62.60 & 75.42 & 81.74 & 83.17 & 75.73
& 46.83 & 72.49 & 86.23 & 88.46 & 73.50
& 25.43 & 35.72 & 45.90 & 48.22 & 38.82 \\
MIM~\cite{10977020}
& 60.53 & 74.20 & 80.30 & 81.97 & 74.25
& 40.56 & 73.59 & 81.87 & 84.72 & 70.19
& 60.58 & 77.03 & 82.29 & 83.98 & 75.97
& 52.46 & 73.18 & 86.74 & 88.19 & 75.14
& 25.92 & 34.58 & 45.57 & 48.06 & 38.53 \\
Hi-End-MAE~\cite{TANG2026103770}
& 69.59 & 78.56 & \underline{82.75} & \textbf{84.53} & \underline{78.86}
& \underline{46.21} & \underline{75.84} & 82.04 & \underline{84.98} & \underline{72.27}
& 73.52 & \underline{79.57} & \underline{82.73} & \underline{84.37} & 80.05
& 63.68 & \underline{76.73} & \underline{88.31} & \underline{90.83} & \underline{79.89}
& 28.73 & 37.14 & 47.13 & 48.99 & 40.50 \\
\midrule
\rowcolor{tablelightblue}
MeDUET$^{\dagger}$
& \underline{70.06} & \underline{78.72} & 82.42 & 83.61 & 78.70
& 45.63 & 75.47 & \underline{82.15} & 84.72 & 71.99
& \underline{75.72} & 79.20 & 81.93 & 83.85 & \underline{80.18}
& \underline{64.83} & 75.90 & 88.09 & 90.62 & 79.86
& \underline{30.80} & \underline{37.82} & \underline{48.29} & \underline{50.05} & \underline{41.74} \\
\rowcolor{tableblue}
\textbf{MeDUET}$^{\ddagger}$
& \textbf{71.98} & \textbf{80.02} & \textbf{83.47} & \underline{84.29} & \textbf{79.94}
& \textbf{48.73} & \textbf{76.38} & \textbf{83.41} & \textbf{85.45} & \textbf{73.49}
& \textbf{76.29} & \textbf{80.89} & \textbf{82.98} & \textbf{84.43} & \textbf{81.15}
& \textbf{67.79} & \textbf{77.36} & \textbf{88.98} & \textbf{91.35} & \textbf{81.37}
& \textbf{32.74} & \textbf{39.17} & \textbf{48.43} & \textbf{50.56} & \textbf{42.73} \\
\bottomrule
\end{tabular}
}
\label{segmentation}
\vspace{-4pt}
\end{table*}

\begin{figure}[t]
  \centering
  \includegraphics[width=1.0\linewidth]{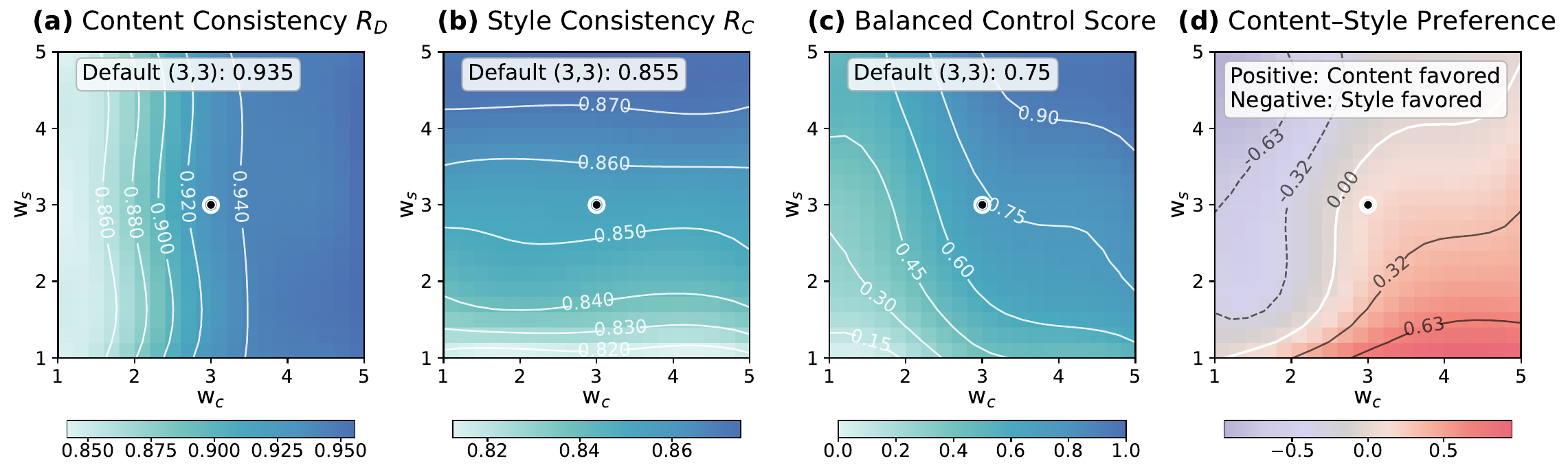}
  \caption{\textbf{Controllability landscape}. (a) Content consistency $R_D$. (b) Style consistency $R_C$. (c) Balanced control score. (d) Content--style preference map, where positive values indicate content-favored control and negative values indicate style-favored control.}
  \label{consistency}
  \vspace{-4pt}
\end{figure}

\begin{table}[t]
\centering
\caption{Results on the classification task on the CC-CCII dataset. $\dagger$: without style augmentation. $\ddagger$: using style augmentation with learned style codes.}
\setlength{\tabcolsep}{4.0mm}
\resizebox{0.80\linewidth}{!}{
\begin{tabular}{lcccc}
\toprule
\multirow{2}{*}{\textbf{Method}} & \multicolumn{4}{c}{\textbf{Accuracy (\%) $\uparrow$}} \\
\cmidrule(lr){2-5}
 & 10\% & 50\% & 100\% & Avg\\
\midrule
\rowcolor{tablegray}
\multicolumn{5}{c}{\textbf{\textit{From Scratch}}} \\
UNETR~\cite{Hatamizadeh_2022_WACV} & 73.80 & 82.40 & 88.65 & 81.62 \\
Swin-UNETR~\cite{hatamizadeh2021swin} & 77.01 & 86.62 & 88.32 & 83.98 \\
\midrule
\rowcolor{tablegray}
\multicolumn{5}{c}{\textbf{\textit{Medical Image SSL}}} \\
Swin-UNETR~\cite{hatamizadeh2021swin} & 77.64 & 87.33 & 89.42 & 84.80 \\
SwinMM~\cite{wang2023swinmm} & 81.49 & 91.27 & 92.23 & 88.33 \\
GVSL~\cite{He_2023_CVPR} & 86.16 & 91.08 & 93.26 & 90.17 \\
VoCo~\cite{Wu_2024_CVPR} & 86.85 & \textbf{91.86} & \underline{93.45} & \underline{90.72} \\
GL-MAE~\cite{11004165} & 80.63 & 88.62 & 92.13 & 87.13 \\
MIM~\cite{10977020} & 82.57 & 89.74 & 92.75 & 88.35 \\
Hi-End-MAE~\cite{TANG2026103770} & 78.76 & 88.15 & 92.59 & 86.50 \\
\midrule
\rowcolor{tablelightblue}
MeDUET$^{\dagger}$ & \underline{88.06} & 90.93 & 93.12 & 90.70 \\
\rowcolor{tableblue}
\textbf{MeDUET}$^{\ddagger}$ & \textbf{88.68} & \underline{91.79} & \textbf{93.59} & \textbf{91.35} \\
\bottomrule
\end{tabular}
}
\label{classification}
% \vspace{-4pt}
\end{table}

\subsubsection{Content and Style Consistency}
To investigate controllable generation, we use dual conditional CFG~\cite{ho2022classifier}, independently scaling content $w_c$ and style $w_s$ during sampling. For content consistency, we fix TotalSegmentator~\cite{wasserthal2023totalsegmentator} content references, sample styles from other domains, sweep $(w_c, w_s)$, and compute the Dice ratio $R_D$ between the segmentations of the references and the generated volumes. For style consistency, we fix style references and report the style-classifier hit rate $R_C$. We use nnU-Net~\cite{isensee2018nnu} as the segmenter and MeDUET's domain classifier as the style classifier. As shown in Fig.~\ref{consistency}, $R_D$ increases mainly with $w_c$ and is only weakly affected by $w_s$, while $R_C$ increases mainly with $w_s$ and remains largely insensitive to $w_c$. The balanced control score becomes higher when both guidance scales are sufficiently strong, and the preference map further reveals a clear transition from style-favored to content-favored regions. 
% The default setting $(w_c, w_s) = (3,3)$ lies in a balanced high-consistency region, demonstrating that MeDUET enables stable and disentangled control over anatomical structure and appearance.

\subsection{Medical Image Analysis}
\label{sec4_3}
We validate MeDUET with both General SSL~\cite{He_2022_CVPR, Chen_2023_WACV, Xie_2022_CVPR, He_2020_CVPR} and Medical SSL~\cite{Hatamizadeh_2022_WACV, hatamizadeh2021swin, wang2023swinmm, He_2023_CVPR, Wu_2024_CVPR, 11004165, 10977020, TANG2026103770}. For medical SSL methods, we either use official pretrained weights or report results drawn from previous work~\cite{Wu_2024_CVPR, TANG2026103770}.

\subsubsection{Performance under Cross-domain and Low-label Settings}
The segmentation results covering \modCT{}, \modMRI{}, and \modPET{} datasets under different labeled data ratios are summarized in Table~\ref{segmentation}. As shown, MeDUET consistently achieves competitive Dice scores, demonstrating stronger generalization than competing SSL methods. In contrast, prior approaches show less stable performance, particularly in low-label regimes and under domain shifts, where their representations are more easily affected by appearance variations. Moreover, adding our style augmentation scheme further improves MeDUET, highlighting the benefit of style-aware latent augmentation under both cross-domain and limited-label settings. This improvement is consistent with the view that MeDUET captures more robust and transferable anatomical semantics with reduced dependence on domain-specific style. Thus, the proposed framework not only improves cross-domain generalization, but also exhibits clear advantages in data efficiency.

\begin{figure}[t]
  \centering
  \includegraphics[width=1.0\linewidth]{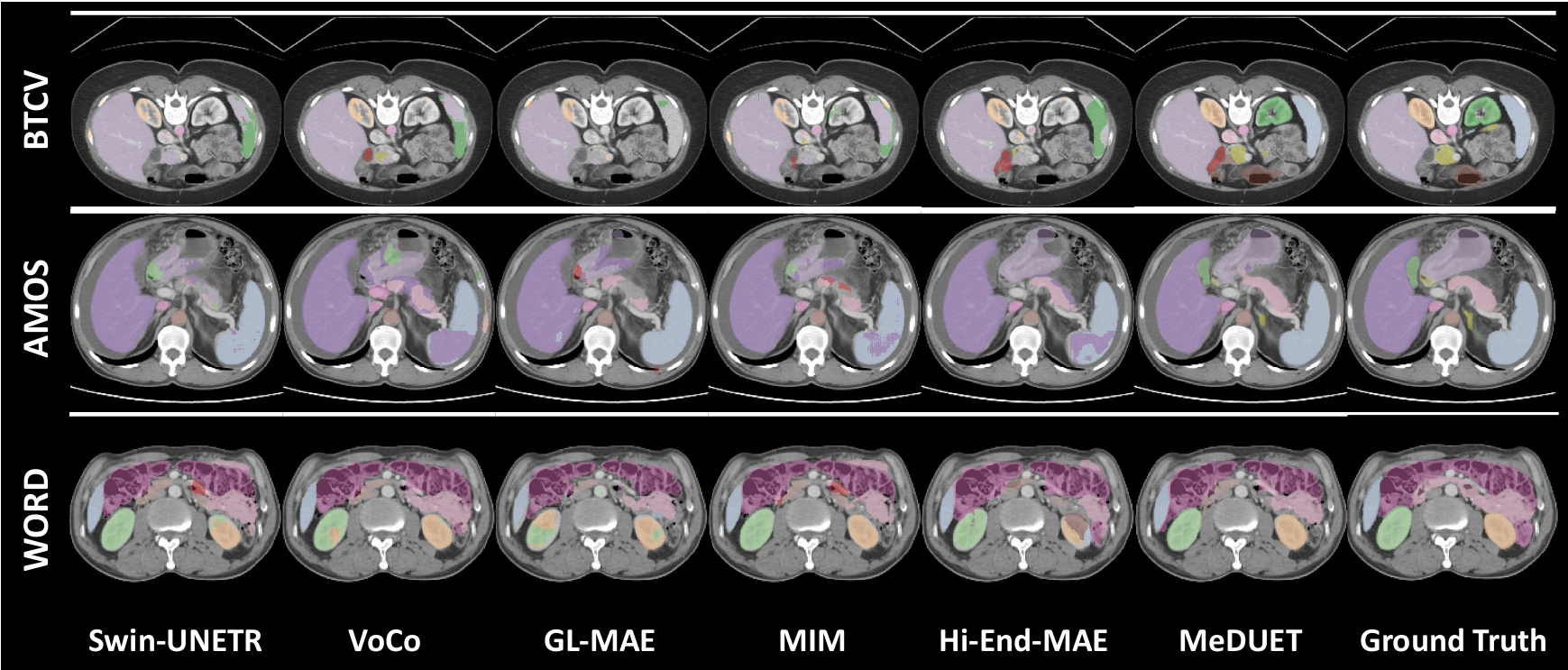}
  \caption{\textbf{Qualitative visualization of one-shot segmentation results} for BTCV (row 1), AMOS (row 2), and WORD (row 3). }
  \label{visual_seg}
  % \vspace{-4pt}
\end{figure}

\begin{figure}[t]
  \centering
  \includegraphics[width=0.90\linewidth]{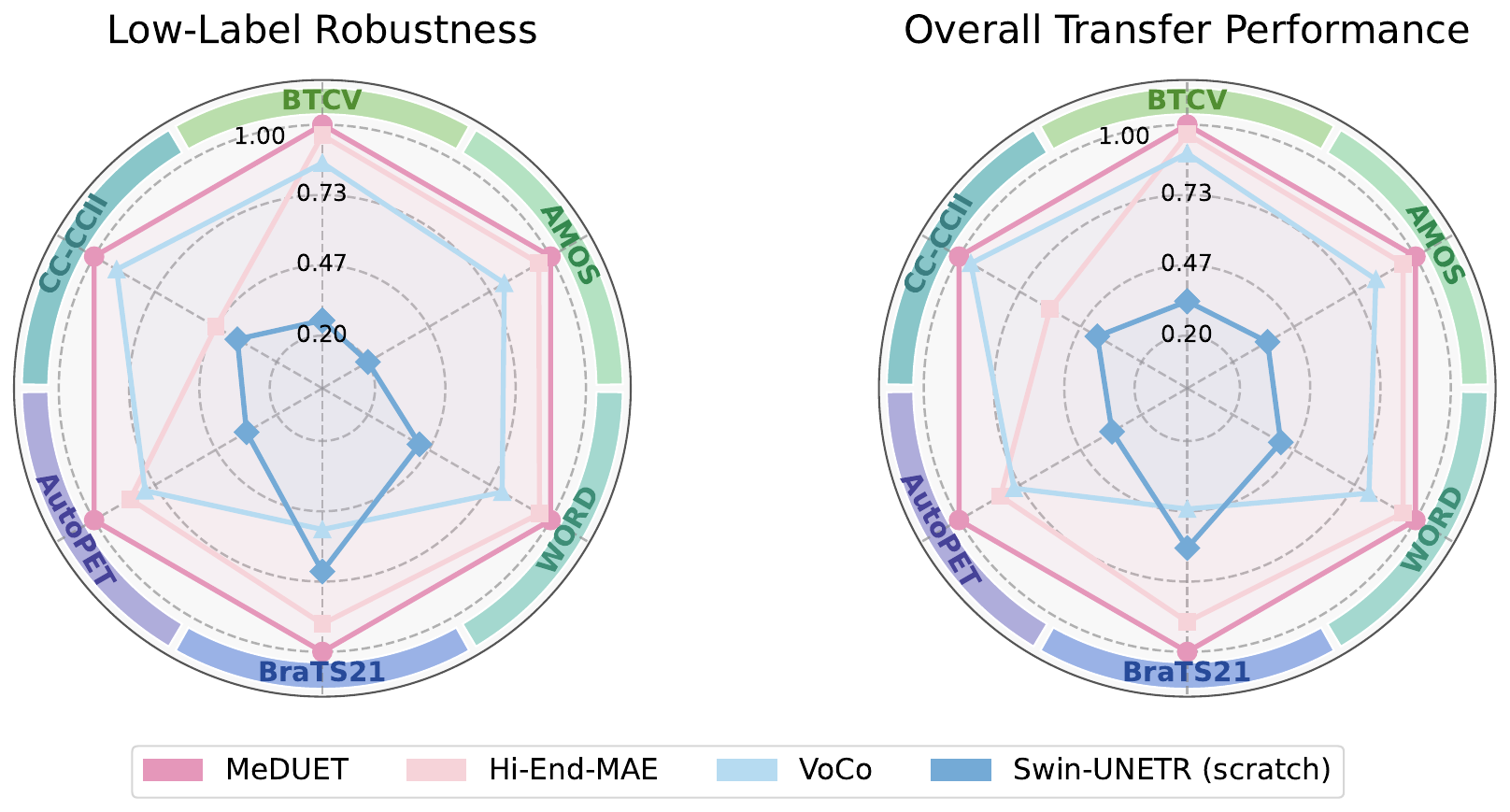}
  \caption{\textbf{Overall comparisons of analysis performance.} \emph{Left}: Low-label robustness under the one-shot setting, together with CC-CCII under the 10\% setting. \emph{Right}: Overall transfer performance summarized by the average performance. Each axis is independently normalized using the full value range of all methods in the corresponding result table and then rescaled to [0.2, 1.0]. }
  \label{overall_seg}
  % \vspace{-4pt}
\end{figure}

\begin{table}[t]
\centering
\setlength{\tabcolsep}{2.0mm}
\caption{Ablation results on loss functions and model components. w/o Disent.: Factor disentanglement is removed by excluding $\mathcal{L}_{d}$ and $\mathcal{L}_{\mathrm{SiQC}}$, and MFTD is modified to perform token-level distillation. w/o Demix.: The demixing module is disabled, resulting in a vanilla masked reconstruction paradigm. }
\resizebox{1.0\linewidth}{!}{
\begin{tabular}{lccc|cc|cc}
\toprule
\multirow{4}{*}{\textbf{Method}} & \multicolumn{3}{c|}{\textbf{Synthesis}} & \multicolumn{4}{c}{\textbf{Analysis}} \\
\cmidrule(lr){2-4}\cmidrule(lr){5-8}
 & \multirow{3}{*}{\textbf{FID} $\downarrow$} & \multirow{3}{*}{\textbf{MMD} $\downarrow$} & \multirow{3}{*}{\textbf{MS-SSIM} $\downarrow$} & \multicolumn{4}{c}{\textbf{Dice Score (\%)$\uparrow$}} \\
\cmidrule(lr){5-8}
 &  &  &  & \multicolumn{2}{c|}{\cellcolor{ctbg}{\color{cttxt}\textbf{BTCV}}} & \multicolumn{2}{c}{\cellcolor{ctbg}{\color{cttxt}\textbf{WORD}}} \\
\cmidrule(lr){5-6}\cmidrule(lr){7-8}
 &  &  &  & 1-shot & 100\% & 1-shot & 100\% \\
\midrule
\rowcolor{tablegray}
\multicolumn{8}{c}{\textbf{\textit{Loss Functions}}} \\
w/o $\mathcal{L}_{d}$             & 0.8147 & 0.5836 & 0.1758 & 64.58 & 82.41 & 60.83 & 82.98 \\
w/o $\mathcal{L}_{\mathrm{MFTD}}$ & 0.7916 & 0.5623 & 0.1689 & 70.13 & 83.84 & 75.09 & 83.76 \\
w/o $\mathcal{L}_{\mathrm{SiQC}}$ & 0.7929 & 0.5619 & 0.1696 & 71.26 & 84.17 & 74.86 & 83.89 \\
\midrule
\rowcolor{tablegray}
\multicolumn{8}{c}{\textbf{\textit{Components}}} \\
w/o Disent. & 0.8460 & 0.5924 & 0.1802 & 62.23 & 83.02 & 62.87 & 82.71 \\
w/o Demix.  & 0.7962 & 0.5677 & 0.1713 & 68.74 & 83.72 & 71.43 & 83.52 \\
\rowcolor{tableblue}
\textbf{MeDUET} & \textbf{0.7874} & \textbf{0.5598} & \textbf{0.1659} & \textbf{71.98} & \textbf{84.29} & \textbf{76.29} & \textbf{84.43} \\
\bottomrule
\end{tabular}
}
\label{ablation}
\end{table}

\subsubsection{One-shot Segmentation Results}
As shown in Table~\ref{segmentation}, under the highly challenging one-shot setting, MeDUET surpasses SSL baselines across 3D medical datasets. This improvement is particularly encouraging because one-shot learning places much stronger demands on the quality and transferability of pretrained representations than standard fine-tuning with sufficient annotations. The consistent advantage of MeDUET suggests that its factor disentanglement mechanism facilitates the model to capture more robust anatomical semantics while suppressing domain-specific appearance bias. Qualitative visualization in Fig.~\ref{visual_seg} further shows that MeDUET produces masks that are visually closer to the ground truth, with more complete structures and cleaner boundaries.

\subsubsection{Generalization Capability for Unseen Modalities}
MeDUET with style augmentation consistently outperforms existing methods on the BraTS 21 and AutoPET III datasets, with the results summarized in Table~\ref{segmentation}. Since BraTS 21 and AutoPET III are unseen \modMRI{} and \modPET{} datasets, respectively, while pretraining is mainly conducted on CT data, this evaluation reflects a particularly challenging unseen-modality transfer setting. The impressive performance of MeDUET therefore confirms that the proposed framework can capture more modality-robust anatomical information. This transferability further highlights the benefit of explicitly disentangling content and style for robust medical image analysis.

\subsubsection{Classification Tasks}
From Table~\ref{classification}, MeDUET achieves competitive performance on the CC-CCII classification task, with consistent gains across different data ratios. Although the gain is relatively modest, its consistency across multiple data settings suggests that the benefit of MeDUET extends beyond segmentation to global semantic recognition tasks.

\subsubsection{Overall Comparisons}
Fig.~\ref{overall_seg} presents comparisons of downstream analysis performance under both low-label and overall-transfer settings. MeDUET exhibits the largest and most balanced profile in both scenarios.
% MeDUET maintains satisfactory performance under extremely limited supervision, while it also preserves strong overall transfer when results are aggregated across the full spectrum of evaluation settings.

\begin{table}[t]
\centering
\setlength{\tabcolsep}{1.0mm}
\renewcommand{\arraystretch}{1.1}
\caption{Ablation experiments on the role of VAE.}
\resizebox{1.00\linewidth}{!}{
\begin{tabular}{lccc|cc|cc}
\toprule
\multirow{4}{*}{\textbf{Method}} & \multicolumn{3}{c|}{\textbf{Synthesis}} & \multicolumn{4}{c}{\textbf{Analysis}} \\
\cmidrule(lr){2-4}\cmidrule(lr){5-8}
 & \multirow{3}{*}{\textbf{FID} $\downarrow$} & \multirow{3}{*}{\textbf{MMD} $\downarrow$} & \multirow{3}{*}{\textbf{MS-SSIM} $\downarrow$} & \multicolumn{4}{c}{\textbf{Dice Score (\%)$\uparrow$}} \\
\cmidrule(lr){5-8}
 &  &  &  & \multicolumn{2}{c|}{\cellcolor{ctbg}{\color{cttxt}\textbf{BTCV}}} & \multicolumn{2}{c}{\cellcolor{ctbg}{\color{cttxt}\textbf{WORD}}} \\
\cmidrule(lr){5-6}\cmidrule(lr){7-8}
 &  &  &  & 1-shot & 100\% & 1-shot & 100\% \\
\midrule
Patch-Volume Autoencoder~\cite{11063450} & 0.7948 & 0.5630 & 0.1686 & 70.37 & 84.12 & 75.13 & \textbf{84.50} \\
MAISI-VAE~\cite{10943915} & \textbf{0.7874} & \textbf{0.5598} & \textbf{0.1659} & \textbf{71.98} & \textbf{84.29} & \textbf{76.29} & 84.43 \\
\bottomrule
\end{tabular}
}
\label{VAE}
% \vspace{-10pt}
\end{table}

\begin{table}[t]
\centering
% \footnotesize
\setlength{\tabcolsep}{2.0mm}
\caption{Linear probe classification results of content and style on the pretraining datasets. We report the Acc (\%) and AUC (\%).}
\resizebox{0.85\linewidth}{!}{
\begin{tabular}{lcc|cc}
\toprule
\multirow{2}{*}{\textbf{Method}} & \multicolumn{2}{c|}{\textbf{Content}} & \multicolumn{2}{c}{\textbf{Style}} \\
\cmidrule(lr){2-3}\cmidrule(lr){4-5}
& \textbf{Acc (\%) $\downarrow$} & \textbf{AUC (\%) $\downarrow$}
& \textbf{Acc (\%) $\uparrow$} & \textbf{AUC (\%) $\uparrow$} \\
\midrule
w/o $\mathcal{L}_{d}$              & 57.53 & 75.48 & 51.46 & 71.52 \\
w/o $\mathcal{L}_{\mathrm{MFTD}}$  & 24.20 & 63.21 & 96.07 & 93.11 \\
w/o $\mathcal{L}_{\mathrm{SiQC}}$  & 28.77 & 68.98 & 92.04 & 90.23 \\
w/o Demix.                         & 23.95 & 62.92 & 96.18 & 93.26 \\
\rowcolor{tableblue}
\textbf{MeDUET}                    & \textbf{23.59} & \textbf{61.56} & \textbf{96.43} & \textbf{93.59} \\
\bottomrule
\end{tabular}
}
\label{linear}
\end{table}

\begin{table*}[t]
\centering
\setlength{\tabcolsep}{2.0mm}
\caption{Linear probe classification results of content and style on the OOD datasets. We report the Acc (\%) and AUC (\%).}
\resizebox{1.00\linewidth}{!}{
\begin{tabular}{lcc|cc|cc|cc|cc|cc}
\toprule
\multirow{3}{*}{\textbf{Method}} 
& \multicolumn{4}{c|}{\cellcolor{ctbg}{\color{cttxt}\textbf{AMOS}}} 
& \multicolumn{4}{c|}{\cellcolor{mribg}{\color{mritxt}\textbf{BraTS 21}}}
& \multicolumn{4}{c}{\cellcolor{petbg}{\color{pettxt}\textbf{AutoPET III}}} \\ 
\cmidrule(lr){2-5}\cmidrule(lr){6-9}\cmidrule(lr){10-13}
& \multicolumn{2}{c|}{\textbf{Content}} 
& \multicolumn{2}{c|}{\textbf{Style}} 
& \multicolumn{2}{c|}{\textbf{Content}} 
& \multicolumn{2}{c|}{\textbf{Style}}
& \multicolumn{2}{c|}{\textbf{Content}} 
& \multicolumn{2}{c}{\textbf{Style}} \\
\cmidrule(lr){2-3}\cmidrule(lr){4-5}
\cmidrule(lr){6-7}\cmidrule(lr){8-9}
\cmidrule(lr){10-11}\cmidrule(lr){12-13}
& \textbf{Acc (\%) $\uparrow$} & \textbf{AUC (\%) $\uparrow$}
& \textbf{Acc (\%) $\downarrow$} & \textbf{AUC (\%) $\downarrow$}
& \textbf{Acc (\%) $\uparrow$} & \textbf{AUC (\%) $\uparrow$}
& \textbf{Acc (\%) $\downarrow$} & \textbf{AUC (\%) $\downarrow$}
& \textbf{Acc (\%) $\uparrow$} & \textbf{AUC (\%) $\uparrow$}
& \textbf{Acc (\%) $\downarrow$} & \textbf{AUC (\%) $\downarrow$} \\
\midrule
% w/o $\mathcal{L}_{d}$              
%   & 75.14 & 81.91 & 79.39 & 83.12 
%   &  &  &  &  
%   &  &  &  &  \\
w/o $\mathcal{L}_{\mathrm{MFTD}}$  
  & 87.25 & 85.32 & 57.04 & 72.63 
  & 91.21 & 90.08 & 74.47 & 85.66
  & 82.86 & 83.15 & 69.53 & 76.06 \\
w/o $\mathcal{L}_{\mathrm{SiQC}}$  
  & 85.43 & 84.18 & 64.47 & 75.79 
  & 89.57 & 88.12 & 80.22 & 86.50
  & 80.95 & 81.73 & 72.17 & 78.53 \\
% w/o Demix.                         
%   &  &  &  &  
%   &  &  &  &  
%   &  &  &  &  \\
\rowcolor{tableblue}
\textbf{MeDUET}                    
  & \textbf{93.31} & \textbf{92.69} & \textbf{42.76} & \textbf{67.18} 
  & \textbf{96.07} & \textbf{95.32} & \textbf{56.30} & \textbf{73.86}
  & \textbf{86.24} & \textbf{87.75} & \textbf{61.27} & \textbf{68.45} \\
\bottomrule
\end{tabular}
}
\label{linear_OOD}
\end{table*}

\subsection{Ablation Study}
We ablate the proposed loss functions, key architectural components, and the VAE tokenizer, with additional hyperparameter analysis reported in the supplementary material.
\subsubsection{Loss Functions and Components}
Table~\ref{ablation} summarizes the contribution of each loss and model component. Consistent with the failure-mode-driven design in Table~\ref{tab:failure_mode_design}, removing any single loss consistently degrades performance, where eliminating the disentanglement design leads to the largest drop, highlighting its importance for medical image analysis. When all losses and components are jointly enabled, MeDUET achieves the best synthesis quality and the highest segmentation Dice, emphasizing that these designs are complementary and beneficial for both generative fidelity and segmentation generalization. 
Beyond these metrics, Sec.~\ref{sec:dis_ana} provides direct factor-diagnostic evidence that MFTD and SiQC improve content/style specialization, as reflected by stronger anatomy retention and reduced factor leakage. This indicates that their effects are more directly captured by empirical factor identifiability than by Dice gains alone.

% \begin{table}[t]
% \centering
% \setlength{\tabcolsep}{1mm}
% \caption{\textbf{Design rationale of MeDUET components.} Each component targets a distinct failure mode of empirical factor identifiability.}
% \resizebox{0.98\linewidth}{!}{
% \begin{tabular}{lccc}
% \toprule
% \textbf{Component} & \textbf{Targeted failure mode} & \textbf{Expected effect} & \textbf{Evidence} \\
% \midrule
% Demixing & Ambiguous factor assignment & Explicit source supervision & Ablation, Fig.~2 \\
% MFTD & Mixed-region leakage & Source-faithful factors & Table~VIII, Fig.~9 \\
% SiQC & Poor factor-space organization & Invariance/discriminability & Table~VIII, Fig.~9--11 \\
% Dual cond. diffusion & Entangled synthesis control & Factor-conditioned generation & Table~I, Fig.~4 \\
% Style aug. & Style overfitting in analysis & Domain-robust readout & Table~III--V \\
% \bottomrule
% \end{tabular}}
% \label{tab:component_rationale}
% \end{table}

\subsubsection{Role of the VAE Tokenizer} 
Given that tailored 3D VAEs for medical imaging remain relatively underexplored, we adopt the Patch-Volume Autoencoder from 3D MedDiffusion~\cite{11063450} for comparison. Table~\ref{VAE} reflects that replacing MAISI-VAE~\cite{10943915} with the Patch-Volume Autoencoder leads to only minor fluctuations, suggesting that the improvements of MeDUET arise primarily from the disentangled pretraining scheme rather than from any specific VAE design.

\begin{figure}[t]
  \centering
  \includegraphics[width=1.00\linewidth]{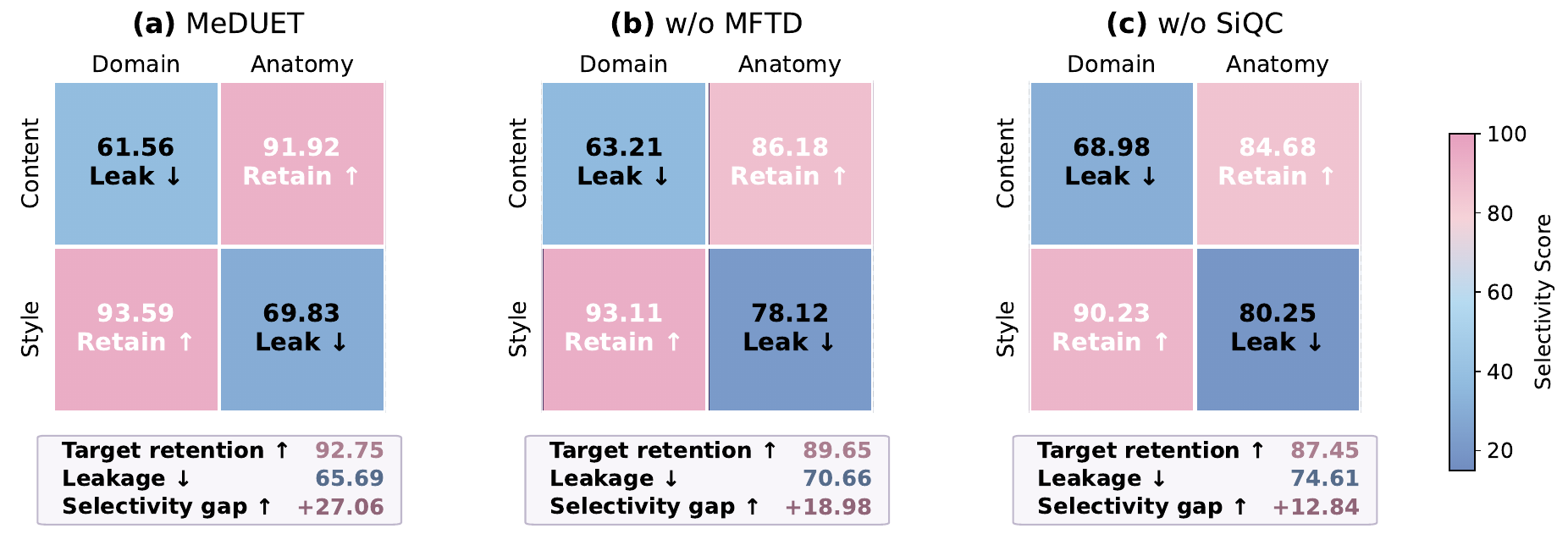}
\caption{\textbf{Factor leakage/retention matrix.} Linear-probe AUC (\%) of domain and anatomy information in content/style factors. Anatomy is averaged over AMOS, BraTS 21, and AutoPET III. The bottom summaries report target retention, i.e., the mean AUC of desired cells (Content--Anatomy and Style--Domain), leakage, i.e., the mean AUC of undesired cells (Content--Domain and Style--Anatomy), and their difference as the selectivity gap.
  % Our MeDUET yields the clearest factor specialization compared with removing MFTD or SiQC.
  }
  \label{factor_matrix}
\end{figure}

\begin{figure}[t]
  \centering
  \includegraphics[width=1.0\linewidth]{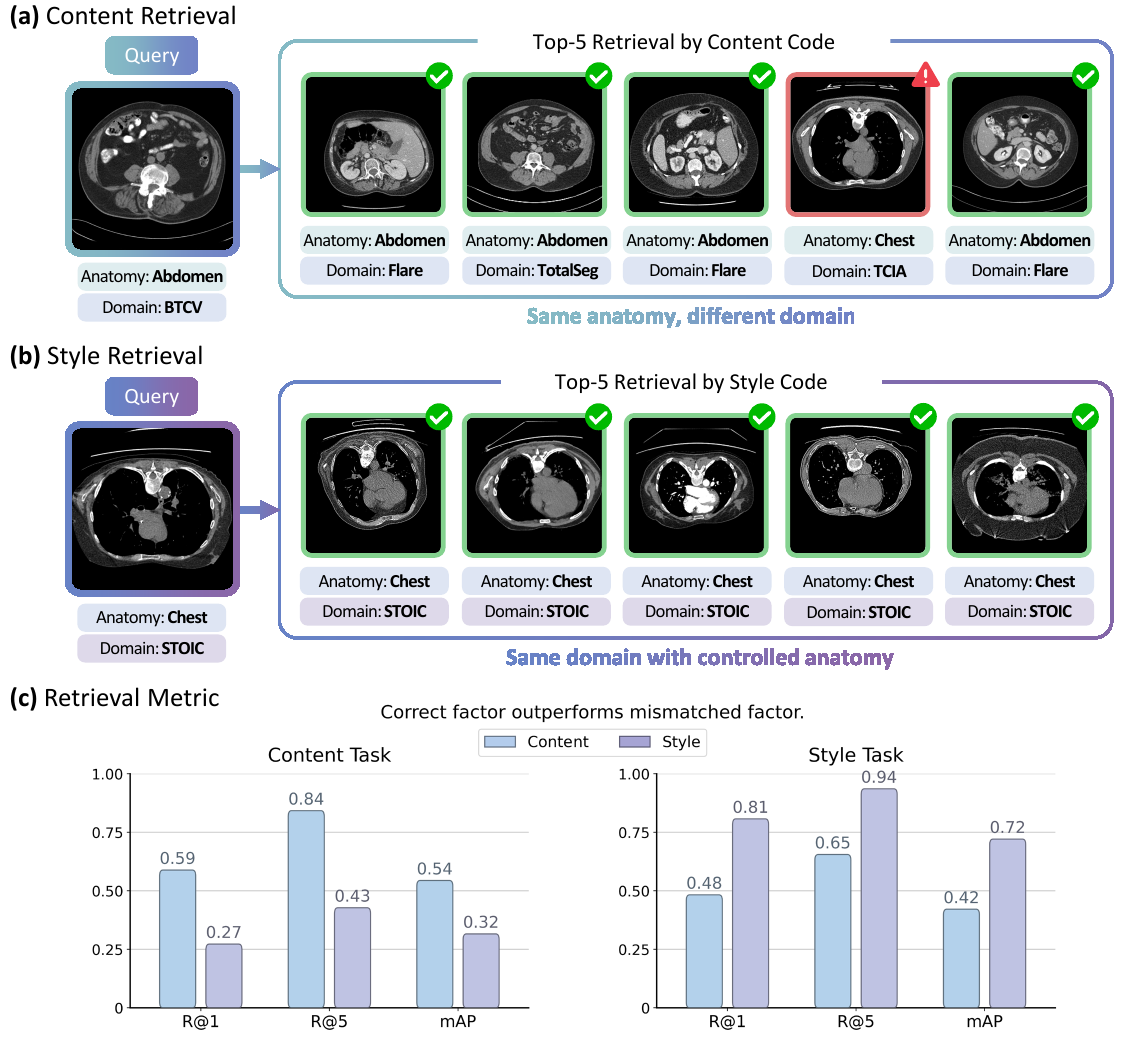}
  \caption{\textbf{Cross-factor retrieval analysis}. Content-code retrieval returns anatomically similar samples across different domains, whereas style-code retrieval retrieves same-domain samples under anatomy control. The quantitative results (R@1, R@5, and mAP) show that the matched factor consistently outperforms the mismatched factor on its corresponding retrieval task.}
  \label{factor_retrieval}
  % \vspace{-4pt}
\end{figure}

\subsection{Disentanglement Analysis}
\label{sec:dis_ana}
Beyond controllable generation and style-aware augmentation, we evaluate whether MeDUET learns identifiable factors by exploring factor specialization, selectivity, latent organization, and robustness to controlled domain shifts. 
% While controllable generation and style-aware augmentation have already been demonstrated in Sec.~\ref{sec4_2} and~\ref{sec4_3}, the experiments here focus on providing direct empirical evidence that each proposed remedy strategy produces the intended effect on factor identifiability.

\subsubsection{Factor Diagnostics}
\label{linear_probe}
% \noindent\textbf{Linear Probe with Domain-category Labels.}
We first assess factor identifiability using linear-probe classification with domain categories as labels. Ideally, the content vector should be domain-invariant, while the style vector should retain domain-specific variations. As shown in Table~\ref{linear}, probes on style vectors achieve much higher domain classification performance, whereas probes on content vectors remain much less discriminative. This separation aligns with the goal of MeDUET, suggesting that acquisition-related cues are captured more strongly in style while the content branch remains less affected by domain.

% \noindent\textbf{OOD Linear Probe with Anatomical-category Labels.}
We further evaluate the disentangled factors under distribution shift using OOD linear probes with anatomical-category labels on AMOS, BraTS 21, and AutoPET III. Table~\ref{linear_OOD} shows that MeDUET achieves the probing pattern that best matches the intended factor roles, indicating that content preserves stronger anatomical semantics, while style remains less informative for anatomy. Fig.~\ref{factor_matrix} summarizes the linear probe results as a factor leakage/retention matrix, where each cell shows the raw probe AUC. The full MeDUET shows the clearest pattern, with content retaining anatomy while suppressing domain information, and style preserving domain cues while remaining less informative for anatomy. 

\subsubsection{Cross-factor Retrieval}
Beyond linear probing, we examine the local neighborhood structure of the factor spaces via retrieval. For content retrieval, we use the content code as the query embedding and retrieve nearest neighbors from different domains, where positives are samples sharing the same anatomy. For style retrieval, we restrict the gallery to the same anatomical group and retrieve neighbors using the style code, where positives are samples from the same domain. 
As illustrated in Fig.~\ref{factor_retrieval}, content codes preferentially retrieve anatomically consistent samples across domains, whereas style codes better recover same-domain neighbors under controlled anatomy. The matched factor outperforms the mismatched factor in R@1, R@5, and mAP, demonstrating that anatomy-dominant information is concentrated in the content space, while style information is captured by the style space.

\subsubsection{Factor Visualization}
\begin{figure}[t]
    \centering
    \includegraphics[width=1.00\linewidth]{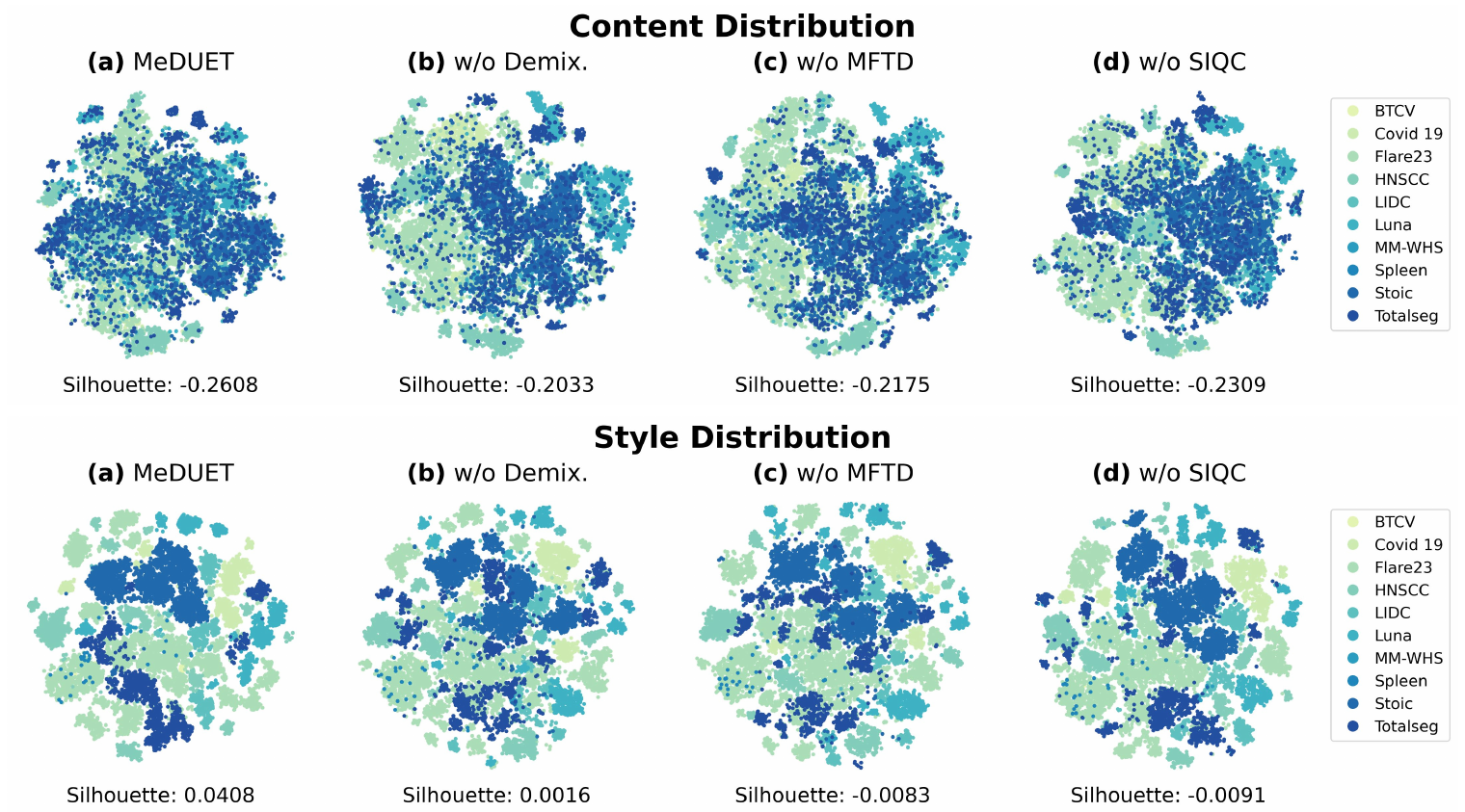}
    \caption{The \textbf{t-SNE visualization} of MeDUET content and style codes, where points are color-coded according to their domain labels.}
    \label{Factor_t-sne}
    % \vspace{-4pt}
\end{figure}

We visualize the t-SNE of factor codes in Fig.~\ref{Factor_t-sne}, where the content vectors of MeDUET are more intermixed across domain labels and yield a lower silhouette score, indicating stronger domain invariance. In contrast, the style vectors show clearer domain separation, suggesting that they capture domain-specific information more effectively. These qualitative results further justify the rationale behind our devised module in promoting better factor separation. 
% To examine generalization, we also visualize factors extracted from the BraTS 21 dataset~\cite{baid2021rsna}, as shown in Fig.~\ref{distribution_MRI}. On this OOD MRI dataset, style embeddings form clear clusters corresponding to the four MRI modalities, while content embeddings remain mixed across modalities. This pattern suggests that the style branch adapts to unseen modality-specific appearance factors, whereas the content branch remains relatively modality-invariant. Together, these results provide empirical evidence that our disentanglement scheme generalizes beyond the training domains.

% \begin{figure}[t]
%     \centering
%     \includegraphics[width=0.90\linewidth]{fig/Factor_visualization_MRI.pdf}
%     \vspace{-10pt}
%     \caption{The t-SNE visualization of MeDUET content and style representations on the BraTS 21 dataset, where points are color-coded according to their modality labels. Left: Content code distribution. Right: Style code distribution.}
%     \label{distribution_MRI}
%     \vspace{-10pt}
% \end{figure}

\begin{table}[t]
\centering
\setlength{\tabcolsep}{1.5mm}
\caption{Anatomy controlled subdomain study using frozen linear probes. Across all three controlled settings, MeDUET shows the clearest factor specialization, with style dominating dataset prediction and content dominating anatomy prediction when anatomy distribution is balanced.}
\label{anatomy_controlled_subdomain}
\resizebox{1.0\linewidth}{!}{
\begin{tabular}{llcc|cc}
\toprule
\multirow{2}{*}{\textbf{Setting}} & \multirow{2}{*}{\textbf{Method}} & \multicolumn{2}{c|}{\textbf{Content}} & \multicolumn{2}{c}{\textbf{Style}} \\
\cmidrule(lr){3-4} \cmidrule(l){5-6}
& & \textbf{Acc (\%)} & \textbf{AUC (\%)} & \textbf{Acc (\%)} & \textbf{AUC (\%)} \\
\midrule

\multirow{3}{*}{\makecell[l]{\textbf{Chest Only Domain} \\ (Style $>$ Content)}}
& w/o $\mathcal{L}_{\mathrm{MFTD}}$  & \textbf{45.09} & \textbf{60.80} & 86.84 & 90.08 \\
& w/o $\mathcal{L}_{\mathrm{SiQC}}$  & 52.14 & 65.07 & 83.10 & 88.75 \\
& \cellcolor{tableblue} \textbf{MeDUET}
& \cellcolor{tableblue} 47.83
& \cellcolor{tableblue} 62.39
& \cellcolor{tableblue} \textbf{87.56}
& \cellcolor{tableblue} \textbf{90.47} \\

\midrule

\multirow{3}{*}{\makecell[l]{\textbf{Balanced Subset Anatomy} \\ (Content $>$ Style)}}
& w/o $\mathcal{L}_{\mathrm{MFTD}}$  & 80.47 & 82.65 & 67.84 & 77.23 \\
& w/o $\mathcal{L}_{\mathrm{SiQC}}$  & 74.91 & 73.58 & 66.04 & 73.67 \\
& \cellcolor{tableblue} \textbf{MeDUET}
& \cellcolor{tableblue} \textbf{83.62}
& \cellcolor{tableblue} \textbf{94.18}
& \cellcolor{tableblue} \textbf{64.53}
& \cellcolor{tableblue} \textbf{71.94} \\

\midrule

\multirow{3}{*}{\makecell[l]{\textbf{Balanced Subset Dataset} \\ (Style $>$ Content)}}
& w/o $\mathcal{L}_{\mathrm{MFTD}}$  & 49.54 & 71.33 & 85.24 & 88.61 \\
& w/o $\mathcal{L}_{\mathrm{SiQC}}$  & 50.78 & 71.64 & 83.80 & 86.59 \\
& \cellcolor{tableblue} \textbf{MeDUET}
& \cellcolor{tableblue} \textbf{43.27}
& \cellcolor{tableblue} \textbf{65.62}
& \cellcolor{tableblue} \textbf{88.46}
& \cellcolor{tableblue} \textbf{90.53} \\

\bottomrule
\end{tabular}
}
\end{table}

\subsubsection{Anatomy Controlled Subdomain Study}
\label{subdomain_study}
To further validate whether the dataset-level domain proxy is confounded with anatomical distribution, we conduct a small-scale anatomy-controlled subdomain study using the same linear probe protocol as in Sec.~\ref{linear_probe}. This control is necessary because our dataset labels may capture not only acquisition-related appearance but also anatomy composition differences, which could otherwise blur the distinction between style-selectivity and anatomy-driven separation.

% \noindent\textbf{Chest-only Controlled Domain Probe.}
We restrict the sample pool to chest datasets only~\cite{clark2013cancer, SETIO20171, doi:10.1148/radiol.2021210384, https://doi.org/10.1118/1.3528204}, fixing anatomy so that the probe target is dataset identity only. Table~\ref{anatomy_controlled_subdomain} demonstrates that style codes still achieve higher domain classification performance than content codes, indicating that style selectivity cannot be explained solely by cross-anatomy differences. We further construct a balanced 2$\times$2 subset using two chest datasets\cite{SETIO20171, doi:10.1148/radiol.2021210384} and two abdomen datasets~\cite{landman2015miccai, ma2024automatic} with equal sample counts, breaking the near one-to-one coupling between dataset and anatomy. Linear probes are trained to predict either anatomy or dataset labels. As reported in Table~\ref{anatomy_controlled_subdomain}, content codes remain more predictive of anatomy while style codes become much less informative for anatomy. These observations confirm that anatomical information is not mainly absorbed by the style branch, and that content remains the primary carrier of anatomical semantics under our dataset-level domain proxy.

\begin{table}[t]
\centering
\setlength{\tabcolsep}{1.5mm}
\caption{Synthesis performance comparison with 3D medical image SSL in the latent space. $\dagger$: using pre-defined metadata vectors. $\ddagger$: using learned content and style vectors.}
\resizebox{1.0\linewidth}{!}{
\begin{tabular}{lccc}
\toprule
\textbf{Method} & \textbf{FID}\ $\downarrow$ & \textbf{MMD}\ $\downarrow$ & \textbf{MS-SSIM}\ $\downarrow$ \\
\midrule
SiT-B/4 + MAE3D$^{\dagger}$~\cite{Chen_2023_WACV} & 0.8416 & 0.5972 & 0.1871 \\
SiT-B/4 + MAE3D$^{\ddagger}$~\cite{Chen_2023_WACV} & 0.8265 \gaingraydown{-0.0151} & 0.5816 \gaingraydown{-0.0156} & 0.1796 \gaingraydown{-0.0075} \\
\midrule
SiT-B/4 + GL-MAE$^{\dagger}$~\cite{11004165} & 0.8322 & 0.5947 & 0.1803 \\
SiT-B/4 + GL-MAE$^{\ddagger}$~\cite{11004165} & 0.8147 \gaingraydown{-0.0175} & 0.5803 \gaingraydown{-0.0144} & 0.1762 \gaingraydown{-0.0041} \\
\midrule
SiT-B/4 + Hi-End-MAE$^{\dagger}$~\cite{TANG2026103770} & 0.8189 & 0.5846 & 0.1773 
\\
SiT-B/4 + Hi-End-MAE$^{\ddagger}$~\cite{TANG2026103770} & 0.8094 \gaingraydown{-0.0095} & 0.5785 \gaingraydown{-0.0061} & 0.1736 \gaingraydown{-0.0037} \\
\midrule
\rowcolor{tablelightblue}
SiT-B/4 + MeDUET$^{\dagger}$  & \underline{0.8011} & \underline{0.5782} & \underline{0.1692} \\
\rowcolor{tableblue}
\textbf{SiT-B/4 + MeDUET$^{\ddagger}$} & \textbf{0.7874 \gainbluedown{-0.0137}} & \textbf{0.5598 \gainbluedown{-0.0184}} & \textbf{0.1659 \gainbluedown{-0.0033}} \\
\bottomrule
\end{tabular}
}
\label{LatentSSL}
\end{table}

\begin{table}[t]
\centering
\caption{Impact of our factors on baseline medical synthesis methods.}
\setlength{\tabcolsep}{2.0mm}
\resizebox{1.0\linewidth}{!}{
\begin{tabular}{lccc}
\toprule
\textbf{Method} & \textbf{FID}\,$\downarrow$ & \textbf{MMD}\,$\downarrow$ & \textbf{MS-SSIM}\,$\downarrow$ \\
\midrule
MAISI~\cite{10943915} & 0.9139 & 0.6292 & 0.2057 \\
\rowcolor{tableblue}
\textbf{\quad + Factor Conditions (Ours)} & 0.9046 \gainbluedown{-0.0093} & 0.6159 \gainbluedown{-0.0133} & 0.1918 \gainbluedown{-0.0139} \\
\midrule
3D MedDiffusion~\cite{11063450} & 0.9216 & 0.6327 & 0.2032 \\
\rowcolor{tableblue}
\textbf{\quad + Factor Conditions (Ours)} & 0.9081 \gainbluedown{-0.0135} & 0.6198 \gainbluedown{-0.0129} & 0.1931 \gainbluedown{-0.0101} \\
\bottomrule
\end{tabular}
}
\label{factor_baseline}
\end{table}

\begin{table}[t]
\centering
\setlength{\tabcolsep}{2.0mm}
\caption{Impact of our plug-and-play style augmentation on baseline medical SSL methods for low-label OOD segmentation.}
\resizebox{1.0\linewidth}{!}{
\begin{tabular}{lcc|cc}
\toprule
\multirow{2}{*}{\textbf{Method}}
& \multicolumn{2}{c|}{\cellcolor{ctbg}{\color{cttxt}\textbf{AMOS}}}
& \multicolumn{2}{c}{\cellcolor{mribg}{\color{mritxt}\textbf{BraTS 21}}} \\
\cmidrule(lr){2-3}\cmidrule(lr){4-5}
& 1-shot & 10\% 
& 1-shot & 10\%\\
\midrule
VoCo~\cite{Wu_2024_CVPR}
& 38.80 & 73.34  
& 49.76 & 71.80   \\
\rowcolor{tableblue}
\textbf{\quad + Style Aug. (Ours)}
& 40.59 \gainblueup{1.79} & 73.58 \gainblueup{0.24}   
& 53.29 \gainblueup{3.53} & 74.92 \gainblueup{3.12}   \\
\midrule
GL-MAE~\cite{11004165}
& 38.17 & 72.43
& 46.83 & 72.49 \\
\rowcolor{tableblue}
\textbf{\quad + Style Aug. (Ours)}
& 39.48 \gainblueup{1.31} & 74.06 \gainblueup{1.63}  
& 49.73 \gainblueup{2.90} & 74.56 \gainblueup{2.07}  \\
\midrule
Hi-End-MAE~\cite{TANG2026103770}
& 46.21 & 75.84  
& 63.68 & 76.73   \\
\rowcolor{tableblue}
\textbf{\quad + Style Aug. (Ours)}
& 48.50 \gainblueup{2.29} & 76.12 \gainblueup{0.28}  
& 65.85 \gainblueup{2.17} & 77.45 \gainblueup{0.72}   \\
\bottomrule
\end{tabular}
}
\label{plug_style_aug}
\end{table}

\begin{table*}[t]
\centering
\caption{The computational cost and wall-clock time comparison of medical SSL methods.}
\setlength{\tabcolsep}{2.0mm}
\resizebox{1.0\linewidth}{!}{
\begin{tabular}{lccccc|ccccc|c}
\toprule
\multirow{2}{*}{\textbf{Method}} & \multirow{2}{*}{\textbf{Backbone}} & \multirow{2}{*}{\textbf{Param. (M)}} & \multirow{2}{*}{\textbf{FLOPs (G)}} &
\multirow{2}{*}{\makecell{\textbf{Per Epoch}\\\textbf{Time (s)}}} &
\multirow{2}{*}{\makecell{\textbf{Pretraining Time}\\\textbf{(GPU hours)}}} &
\multicolumn{5}{c|}{\textbf{Dice Score (\%) $\uparrow$}} &
\textbf{Acc} \textbf{(\%) $\uparrow$} \\
\cmidrule(lr){7-12}
& & & & & & {\cellcolor{ctbg}{\color{cttxt}\textbf{BTCV}}} & {\cellcolor{ctbg}{\color{cttxt}\textbf{AMOS}}} & {\cellcolor{ctbg}{\color{cttxt}\textbf{WORD}}} & {\cellcolor{mribg}{\color{mritxt}\textbf{BraTS 21}}} & {\cellcolor{petbg}{\color{pettxt}\textbf{AutoPET III}}} & {\cellcolor{ctbg}{\color{cttxt}\textbf{CC-CCII}}} \\
\midrule
VoCo~\cite{Wu_2024_CVPR} & Swin-B & 127.44 & 1264.9 & 1867.3 & 208.8 & 76.64 & 69.59 & 77.02 & 74.33 & 39.77 & 90.72 \\
MIM~\cite{10977020} & ViT-B  & 71.0  & 1002.5 & 615.4  & 683.8 & 74.25 & 70.19 & 75.97 & 75.14 & 38.53 & 88.35 \\
Hi-End-MAE~\cite{TANG2026103770} & ViT-B  & 98.9   & 133.9  & 164.6  & 731.6 & 78.86 & 72.27 & 80.05 & 79.89 & 40.50 & 86.50 \\
\midrule
w/o SiQC & ViT-B & 105.1 & 50.1 & 72.0 & 399.8  & 78.50 & 72.79 & 80.53 & 80.79 & 42.47 & 91.14 \\
w/o MFTD & ViT-B & 105.1 & 50.1 & 70.7 & 392.5  & 79.13 & 72.46 & 80.29 & 81.16 & 42.18 & 91.08 \\
\rowcolor{tableblue}
\textbf{MeDUET} & ViT-B & 105.1 & 50.1 & 72.9 & 405.2 &
\textbf{79.94} & \textbf{73.49} & \textbf{81.15} & \textbf{81.37} & \textbf{42.73} & \textbf{91.35} \\
\bottomrule
\end{tabular}
}
\label{computational}
\end{table*}

\subsection{Isolating the Transfer Utility of Disentangled Factors}
To distinguish the benefit of explicit factor disentanglement from that of generic SSL pretraining, we further isolate whether the learned content and style factors provide additional transfer utility under matched latent-space settings.

\subsubsection{Medical SSL for Generative Transfer}
We reimplement representative medical SSL~\cite{Chen_2023_WACV, 11004165, TANG2026103770} within the same VAE latent space and transfer their encoders to initialize SiT. Table~\ref{LatentSSL} illustrates that our MeDUET outperforms baselines, where the introduction of our learned factors leads to varying degrees of improvement for latent medical SSL. Crucially, with setups all held fixed, the consistent gains across three metrics isolate the benefit of our disentangling components rather than model capacity or data advantages.
% Taken together, these results establish MeDUET as a more transferable and effective pretraining paradigm for 3D medical imaging than latent-space SSL alone.

\subsubsection{Factor Impact on Medical Synthesis Baselines}
We replace the original metadata conditions in medical synthesis baselines~\cite{10943915, 11063450} with our content and style vectors. As shown in Table~\ref{factor_baseline}, plugging in our factors yields consistent improvements on all three metrics for both baselines, even though the generative architectures and training setups are unchanged. This pattern supports that MeDUET provides more informative conditioning signals than handcrafted metadata by encoding continuous image-derived anatomy and appearance cues that are not fully captured by discrete labels.

\subsubsection{Plug-and-play Style Augmentation for Medical SSL Analysis}
We further investigate whether our style factors can benefit medical SSL methods for analysis tasks. Table~\ref{plug_style_aug} indicates OOD segmentation improvements for three baselines after incorporating our style augmentation. The gains support that our style space captures transferable domain-sensitive variations that can be injected into downstream training to enhance robustness in low-label and cross-domain scenarios. 
% Nevertheless, MeDUET still achieves the best overall performance, indicating that explicit disentanglement during pretraining is more effective than applying style augmentation only at the downstream stage.

\subsection{Computational Analysis} 
\label{compute}
Although the overall pipeline of MeDUET includes multiple components and objectives, its computational cost remains practical. To verify this, we compare MeDUET with medical SSL methods~\cite{Wu_2024_CVPR, 10977020, TANG2026103770} in Table~\ref{computational}. For fairness, all methods use identical configurations, and the pretraining time is measured on four NVIDIA A100 GPUs. MeDUET achieves the best Dice and Accuracy while also requiring the lowest FLOPs and per-epoch wall-clock time. This efficiency mainly comes from performing SSL in the VAE latent space with a $4 \times$ spatial compression ratio, which substantially reduces the number of tokens. We further compare MeDUET with w/o MFTD and w/o SiQC, where both MFTD and SiQC introduce no extra trainable parameters and only negligible runtime overhead, while consistently improving downstream performance. This makes them computationally efficient additions with a favorable effectiveness to cost trade-off.
\section{Discussion}
\label{sec:discussion}

\subsubsection{Domain Setup}
Not all medical datasets provide reliable fine-grained style metadata, such as scanner vendor, field strength, or reconstruction kernel. We therefore use dataset-level domain IDs as weak domain labels to maintain consistency across datasets. Although this setup is coarse, and dataset identity may partially correlate with anatomical distribution, our experiment (Sec.~\ref{subdomain_study}) provides direct evidence that the learned factorization is not driven by this confounding alone. 
% Specifically, the experimental findings support the practicality of dataset-level domain labels in our setting, while finer-grained or more adaptive domain definitions remain an important direction for future work.

% Not all medical datasets provide reliable fine-grained style metadata, such as scanner vendor, field strength, kVp, or reconstruction kernel. Therefore, instead of relying on such annotations, we use dataset-level domain IDs as weak domain labels to maintain consistency across datasets. Although this setup is coarse and each dataset may still include multiple protocols or gradual appearance shifts, it captures the dominant cross-dataset style differences that the style branch is intended to model. In practice, our experiments show that this simple domain definition is sufficient to yield stable disentanglement and strong downstream performance. Exploring finer-grained or more adaptive domain definitions is an important direction for future work.

\subsubsection{Empirical Factor Identifiability}
We emphasize that the identifiability discussed in this work is empirical rather than theoretical. MeDUET does not provide formal guarantees that content and style factors are uniquely recoverable. Instead, we use identifiability to describe the extent to which the learned factors exhibit stable factor selectivity under controlled diagnostics, including linear probes, retrieval, visualization, and factor-conditioned generation. Under this empirical notion, our results consistently suggest that the learned factor spaces behave in a factorized and practically transferable manner. We argue that this empirical notion is particularly appropriate for 3D medical imaging, where formal identifiability conditions are rarely satisfied in practice due to scanner heterogeneity, protocol variability, and limited sample diversity.

\subsubsection{Design Justification of MeDUET}
A possible concern is that MeDUET introduces several components. As summarized in Table~\ref{tab:failure_mode_design}, these mechanisms are problem-driven, with each targeting a distinct empirical factor-identifiability failure mode and supported by isolated evidence. They are mainly used during pretraining, while the deployed synthesis and analysis models remain comparable to standard latent diffusion and UNETR-style pipelines. Together with the computational analysis in Sec.~\ref{compute}, this indicates that MeDUET provides a principled design with reasonable practical cost.

% \subsubsection{Design Justification of MeDUET}
% A possible concern is that MeDUET contains multiple components. 
% As summarized in Table~\ref{tab:failure_mode_design}, each component corresponds to a distinct empirical factor-identifiability failure mode and is supported by isolated ablation or diagnostic evidence. Moreover, most additional mechanisms are used only during pretraining, so the deployed synthesis and analysis models remain comparable. The computational analysis in Sec.~\ref{compute} shows that MeDUET remains efficient in practice, indicating that the overall design is both principled and computationally reasonable.

\subsubsection{Limitations and Future Directions}
Despite its strong performance, MeDUET still has several limitations. First, although MeDUET maintains standard downstream deployment, its factor-learning pretraining is more resource-intensive than simpler SSL baselines. Future work will explore lightweight factorization designs to reduce this overhead. Second, we primarily evaluate segmentation and classification as downstream analysis tasks. Extending MeDUET to other applications, such as detection, registration, or report generation, remains an important direction for future work.

% Despite its strong performance, MeDUET still has several limitations. First, our generative evaluation focuses on datasets with anatomical distributions similar to those seen during pretraining, while cross-dataset or OOD synthesis deserves further study. Second, the framework introduces multiple modules, making it more complex than simpler SSL baselines. Although operating in the latent space keeps the computational cost manageable, pretraining remains relatively resource-intensive. Finally, we primarily evaluate segmentation and classification as downstream analysis tasks. Extending MeDUET to other applications, such as detection, registration, or report generation, as well as to more heterogeneous real-world multi-institutional settings, remains an important direction for future work.
\section{Conclusion}
\label{sec:conclusion}
In this work, we present MeDUET, a unified pretraining framework for 3D medical imaging that learns empirically disentangled content-style factors within a VAE latent space and transfers seamlessly to both generative and analysis tasks. Our design explicitly enforces disentanglement so that content governs anatomical structure while style captures domain attributes. Extensive experiments demonstrate that MeDUET supports diverse synthesis and analysis applications, achieving strong synthesis performance and competitive or superior analysis performance across diverse medical benchmarks. By unleashing the potential of SSL for both medical image generation and understanding, we believe MeDUET provides valuable insights that can guide future research on pretraining strategies for 3D medical imaging.

% \section*{Acknowledgement}
% The computations described in this research were performed using the Baskerville Tier 2 HPC service. Baskerville was funded by the EPSRC and UKRI through the World Class Labs scheme (EP/T022221/1) and the Digital Research Infrastructure programme (EP/W032244/1) and is operated by Advanced Research Computing at the University of Birmingham.

\bibliographystyle{IEEEtran}
\bibliography{TIP.bib}

% \newpage

 

% \end{IEEEbiography}



\vfill

\end{document}